\documentclass[twocolumn,showpacs,preprintnumbers,superscriptaddress,amsmath,amssymb,nofootinbib, floatfix]{revtex4}
\usepackage{epsfig,graphics,color,graphicx,amsmath}
\usepackage[section]{placeins}
\newcommand{\cd}{\makebox[0.08cm]{$\cdot$}}
\newcommand{\bg}[1]{\mbox{\boldmath $#1$}}
\newcommand{\sla}{\not\!}
\begin{document}
\vspace{0.5cm}
\title{Ab initio nonperturbative calculation of physical observables \\in light-front dynamics.
Application to the Yukawa model}
\author{V.~A.~Karmanov}
\affiliation {Lebedev Physical Institute, Leninsky Prospekt 53,
119991 Moscow, Russia}
\author{J.-F.~Mathiot}
\affiliation {Clermont Universit\'e, Laboratoire de Physique
Corpusculaire, \\ BP10448, F-63000 Clermont-Ferrand, France}
\author{A.~V.~Smirnov}
\affiliation {Lebedev Physical Institute, Leninsky Prospekt 53,
119991 Moscow, Russia}

\bibliographystyle{unsrt}

\begin{abstract}
We present a coherent and operational strategy to calculate, in a
nonperturbative way, physical observables in light-front dynamics.
This strategy is based on the decomposition of the state vector of
any compound system in Fock components, and on the covariant
formulation of light-front dynamics, together with the so-called
Fock sector dependent renormalization scheme. We apply our
approach to the calculation of the electromagnetic form factors of
a fermion in the Yukawa model, in the nontrivial three-body Fock
space truncation, for rather large values of the coupling
constant. We find that, once the renormalization conditions are
properly taken into account, the form factors do not depend on the
regularization scale, when the latter is much larger than the
physical masses. We then extend the Fock space by including
antifermion degrees of freedom.
\end{abstract}
\pacs {11.10.Ef, 11.10.Gh, 11.10.St}
\maketitle

%%%%%%%%%%%%%%%%%%%%%%%%%%%%%%%%%%%%%
\section{Introduction}\label{intro}
We have developed over  the last
 years a
general strategy to calculate physical observables of compound
systems in a nonperturbative
framework~\cite{bckm,kms_04,kms_07,kms_08,kms_10,mstk}. This
strategy is based on light-front dynamics (LFD), a relativistic
Hamiltonian formalism advocated first by Dirac in
1949~\cite{dirac}. In its original formulation, the state vector
of any compound system evolves in the light-front time $\tau =
t+z$, instead of the usual time $t$. Since the physical vacuum in
LFD is identical to the free vacuum (see e.g.
Ref.~\cite{bpp} and references therein), it is
then natural to decompose the state vector of a compound system in
Fock components, since this decomposition does not include any
vacuum fluctuations, but contains physical (asymptotic) states
only. The problem of finding the state vector can thus be
formulated as a $N$-body problem.

For obvious practical reasons, the Fock decomposition should be
truncated to  a finite number of states (Fock sectors). This
truncation should be strictly controled in order to be able to
make reliable predictions for physical observables, order by order
in the Fock expansion. This is made possible due to two important
breakthroughs:
\begin{itemize}
\item The formulation of LFD in a covariant way~\cite{karm76},
which enables the strict control of any violation of rotational
invariance, when the Fock space is truncated. This formulation,
called covariant light-front dynamics (CLFD), has proven to be
very powerful in the description of relativistic properties of
few-body systems~\cite{cdkm}. \item The development of an
appropriate renormalization
procedure --- the so-called Fock sector dependent
renormalization (FSDR) scheme --- which enables to calculate
regularization scale invariant observables order by order in the
Fock expansion~\cite{kms_08}.
\end{itemize}

In the simplest, two-body, Fock space truncation, our formalism is
equivalent to summing the irreducible block --- the fermion
self-energy calculated in the second order of
perturbation theory --- to all orders in the chain approximation.
This
equivalence is caused by the fact that all the
chain type contributions are restricted to the
two-body Fock sector. Such a result is a direct consequence of our
FSDR scheme and the corresponding renormalization conditions.

The first nontrivial calculation corresponds to the three-body
Fock space truncation which incorporates, in the Yukawa model for
instance, fluctuations of the state vector involving one fermion
($f$), one fermion and one boson ($fb$), one fermion and two bosons
($fbb$) Fock sectors. This calculation includes overlapping type
(divergent) diagrams summed to all orders in the coupling constant.

Within the FSDR framework, the first calculation of
a physical observable --- the anomalous magnetic
moment (AMM) of a fermion in the Yukawa model
--- has been done in Ref.~\cite{kms_10}, using the Pauli-Villars
(PV) regularization scheme, as
proposed in Ref.~\cite{PV}. The calculation has
shown nice convergence of the results as a function of the
regularization scale (the PV boson mass in our calculation) for
values of the coupling constant $\alpha \equiv g^2/4\pi$ of order
of 0.2. For stronger coupling, $\alpha\sim 0.5$, some dependence
(though rather weak) of the
AMM on the PV boson mass was detected. While
this range of the coupling constant values is not particularly
small (as compared for instance to the electromagnetic coupling
constant), it shows however that the truncation of the Fock
expansion was not completely under control.

We detail in the present study an extension of
our previous approach~\cite{kms_10}, in order to control, order by
order in the Fock expansion, the regularization scale invariance
of physical observables. Our derivation is based on the full
account of the renormalization conditions, using the FSDR scheme.
We shall see that in the truncated Fock space, the bare coupling
constant and counterterms are no more true constants, but become
naturally dependent on one of the kinematical variables, as
already emphasized in Ref.~\cite{WG}. This dependence is
determined unambiguously by the renormalization conditions and
allows to restore rotational invariance broken by the truncation,
as well as the independence of observables on the
masses of the PV particles, when the latter ones are much larger
than the physical masses. In Ref.~\cite{kms_10} the fact that the
bare coupling constant and counterterms are, {\em a
priori},  functions of kinematical variables was not taken into
account in full measure. The calculations were done for
fixed values of
the kinematical variables.

The kinematical dependence of the bare coupling
constant and counterterms is intimately linked to the
Fock space truncation. Its
explicit form is strongly affected by the Fock space "contents".
Thus, in the Yukawa model considered in the three-body
($f+fb+fbb$) approximation, this dependence is quite sizeable.
It is substantially reduced, when antifermion degrees of freedom (d.o.f.),
namely, the additional three-body Fock sector $ff\bar{f}$,
are taken into account. In leading order of perturbation theory, we show that the inclusion
of the antifermion d.o.f. results in the independence of the bare
coupling constant and counterterms
on the
kinematical variables.

The plan of the article is as follows. We recall in
Sec.~\ref{general} the general properties of our formalism. We
discuss in Sec.~\ref{renor} the renormalization conditions. In
Sec.~\ref{ffbfbb} we obtain a system of renormalized equations for
the Fock components in the Yukawa model within the $f+fb+fbb$ Fock
space truncation and calculate the fermion
electromagnetic form factors. In Sec.~\ref{anti} we extend the
Fock space by the inclusion of the $ff\bar{f}$ Fock sector and
discuss the role of antiparticle d.o.f. We present our conclusions
in Sec.~\ref{conc}. The contribution of antifermion d.o.f. to the
equations for the Fock components is given
in Appendix~\ref{ap1}.

%%%%%%%%%%%%%%%%%%%%%%%%%%%%%%%%%%%%%%%%
\section{General framework} \label{general}
\subsection {Covariant formulation of light-front dynamics}
In the traditional form of LFD,  the state vector of a compound
system is defined on the light-front plane $t+z=0$ (with $c=1$)
rather than on the equal-time plane $t=0$. In order to recover
explicitly rotational invariance, the state
vector is defined, in CLFD, on the light-front plane of general
orientation $\omega \cd x= 0$, where $\omega$ is an arbitrary
four-vector restricted by the condition $\omega^2=0$
\cite{karm76,cdkm}. The traditional form of LFD is recovered by
using $\omega = (1,0,0,-1)$.

The state vector $\phi(p)$ of a particle with
the mass $M$ should satisfy the Poincar\'e group
equations, and among them
\begin{equation}\label{kt15}
\hat{P}^{2} \phi(p)=M^2 \phi(p).
\end{equation}
The momentum operator  $\hat{P}$ is decomposed, on the light
front, in terms of
its free and interaction parts:
\begin{equation}
\label{kt1} \hat{P}_{\rho}=\hat{P}^{(0)}_{\rho}
+\hat{P}^{int}_{\rho},
\end{equation}
where, in terms of the interaction Hamiltonian $H^{int}(x)$,
\begin{equation}
\label{kt4} \hat{P}^{int}_{\rho}=\omega_{\rho} \int H^{int}(x) \delta(\omega\cd x)\ d^4x.
\end{equation}

According to the general properties of LFD, we  decompose the
state vector of a physical system in Fock sectors. We have
schematically
\begin{eqnarray}\label{phi}
\phi(p)&=&
\sum_{n=1}^{ \infty} \int d\tilde{D}_n \ \phi_n(k_1,\ldots,k_n;p) \nonumber \\
&\times& \delta^{(4)}(k_1+\ldots +k_n-p-\omega \tau_n) \left\vert n
\right>,\label{Fock}
\end{eqnarray}
where $\left\vert n \right>$ is the state containing $n$ free
particles with the four-momenta $k_1,\ldots,k_n$ and $\phi_n$'s
are relativistic $n$-body wave functions, the so-called Fock
components. The phase space volume element is represented by
$d\tilde{D}_{n}$. All the four-momenta are on the corresponding
mass shells: $k_i^2=m_i^2$, $p^2=M^2$, $(\omega\tau_n)^2=0$. Note
the peculiar overall four-momentum conservation law given by the
$\delta$-function. It follows from the general transformation
properties of the light-front plane $\omega \cd x=0$ under
four-dimensional translations~\cite{cdkm}. The
scalar quantity $\tau_n$ is a measure of how
far the $n$-body system is off the energy shell (on the energy
shell $\tau_n=0$). It is completely determined by this
conservation law and the on-mass-shell conditions for each
individual particle momentum. We get
\begin{equation}
\label{tau}
2 \omega \cd p \ \tau_n=(s_n-M^2),\mbox{with}\ s_n=(k_1+\ldots +k_n)^2.
\end{equation}
The state $\vert n\rangle$ can be written as
\begin{equation}
\label{freefield}
\left\vert n \right> \equiv d^\dagger (k_1) d^\dagger(k_2) \ldots
d^\dagger(k_{n}) \left\vert 0 \right>,
\end{equation}
where $d^\dagger$  is a generic notation for the fermion and boson
creation operators. To completely determine the state vector, we
normalize it according to
\begin{equation}
\label{normep} \phi(p') ^\dagger \phi(p) = 2 p_0 \delta^{(3)}({\bf
p'}-{\bf p}).
\end{equation}
With the decomposition~(\ref{Fock}), the
normalization condition~(\ref{normep}) writes
\begin{equation} \label{Inor}
 \sum_{n=1}^\infty I_n=1,
\end{equation}
where $I_n$ is the contribution of the $n$-body Fock sector to the
norm. For the particular case of the Yukawa model, an explicit
formula for $I_n$ can be found in Ref.~\cite{kms_08}.

\begin{figure}[ht!]
\includegraphics[width=12pc]{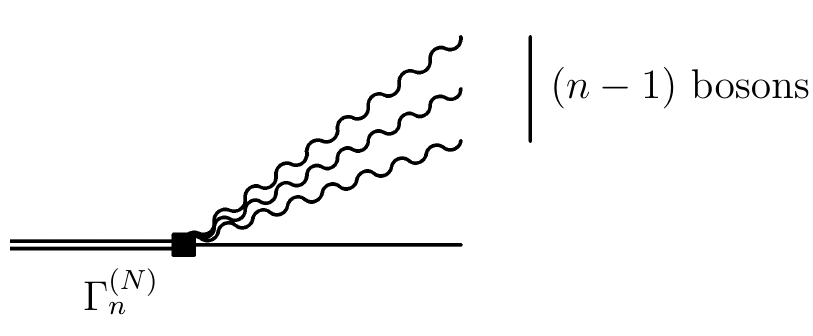}
 \caption{$n$-body vertex function
for the Fock space truncation of order $N$, for a physical fermion
(double straight line) made of a constituent fermion (single
straight line) coupled to $(n-1)$ bosons (wavy lines).}
 \label{gamman}
\end{figure}

We shall concentrate, in the following, on systems composed of a
spin-$1/2$ fermion coupled to scalar bosons. It is convenient to
introduce, instead of the wave functions $\phi_n$, the vertex
functions $\Gamma_n$ (which we will also refer to as Fock
components), defined by
\begin{equation}
\label{Gn} \bar u(k_1) \Gamma_n u(p) =(s_n-M^2)\phi_n \equiv 2
\omega \cd p \ \tau_n \phi_n,
\end{equation}
where $k_1$ is the four-momentum of the constituent fermion. When
the Fock space is truncated to order $N$ [i.~e. in the sums over
$n$ in Eqs.~(\ref{phi}) and~(\ref{Inor}) the terms with $n\leq N$
only are retained], it is necessary to keep track of the order of
truncation in the calculation of the vertex function. For this
purpose, we will denote the latter by $\Gamma_n^{(N)}$, but omit
the superscript $(N)$ when it is not necessary. $\Gamma_n^{(N)}$
is represented graphically by the diagram shown in
Fig.~\ref{gamman}.

It is convenient to introduce the notation
\begin{equation}\label{calG}
{\cal G}(p)=2(\omega\cd p)\hat{\tau}\phi(p),
\end{equation}
where $\hat \tau $ is the operator which, acting on a given
component $\phi_{n}$ of $\phi(p)$, gives $\tau_n \phi_{n}$. ${\cal
G}(p)$ has the Fock decomposition which is obtained from
Eq.~(\ref{Fock}) by the replacement of the wave functions $\phi_{n
}$ by the vertex functions $\Gamma_{n }$. We can thus cast the
eigenstate equation in the form~\cite{bckm}
\begin{equation}\label{eq3}
{\cal G}(p) = \frac{1}{2\pi}\int
\left[-\tilde{H}^{int}(\omega\tau)\right]\frac{d\tau}{\tau} {\cal G}(p),
\end{equation}
where $\tilde{H}^{int}$ is the interaction Hamiltonian in momentum
space. This equation is quite general and equivalent to the
eigenstate equation~(\ref{kt15}). It is nonperturbative.

The graph technique rules derived in Ref.~\cite{cdkm} for the
calculation of $S$-matrix elements in CLFD refer precisely to the
calculation of matrix elements of $-\tilde H^{int}/(2\pi\tau)$. A
system of coupled eigenstate equations for the Fock components of
the state vector can thus be obtained from Eq.~(\ref{eq3}) by
substituting there [via Eq.~(\ref{calG})] the Fock
decomposition~(\ref{Fock}) and calculating the matrix elements of
the operator $-\tilde H^{int}/(2\pi\tau)$ in the Fock space.

Since our formalism is explicitly covariant, the spin structure of
the wave function $\phi_{n}$ is very simple. Indeed, its
construction is of purely kinematical nature~\cite{cdkm}. It
should incorporate however $\omega$-dependent components. The
structure of the two-body components for QED and the Yukawa model
was detailed in Ref.~\cite{kms_04}, while the three-body ($fbb$)
component in the Yukawa model was constructed in
Ref.~\cite{kms_10}. For the purpose of the present study, we
recall here the spin structure of the two-body component in the
Yukawa model:
\begin{equation}
\label{g2}  \bar u(k_1) \Gamma_2 u(p)=\bar{u}(k_1) \left[b_1   +
\frac{M \sla \omega }{\omega \cd p}\ b_2\right]  u(p).
\end{equation}
The coefficients $b_1$ and $b_2$ are scalar functions determined
by dynamics.

%%%%%%%%%%%%%%%%%%%%%%%%%%%%%%%%%%%%%%%%%%%%%%%%%%%%%%%%%
\subsection {Fock sector dependent renormalization scheme}
In standard renormalization theory, the bare parameters (the whole
set of bare coupling constants and counterterms) are determined by
relating them to physically observable quantities. To perform this
strategy in practice, two important questions should be clarified,
when using CLFD.

({\it i}) The explicit form of the relationship between the bare
and physical parameters depends on the approximation used. For
instance, in perturbation theory the bare parameters are
polynomial functions of the physical coupling constant. The term
with the maximal degree of the coupling constant is determined
within a given order of perturbation, while those of lower degrees
are taken unchanged from lower orders. This guarantees that
calculated physical observables are regularization scale invariant
in any order of the perturbative expansion. In our nonperturbative
approach based on the truncated Fock decomposition, an analogous
requirement implies that the bare parameters should depend {\em a
priori} on the Fock sector in which they are considered
\cite{perry}.

({\it ii}) In order to express the bare parameters through the
physical ones, and vice versa, one should be able to calculate
physical observables. In LFD, these ones can not depend on the
choice of the orientation of the light-front plane. Such a
situation indeed takes place, for instance, order by order in
perturbation theory, provided the regularization of divergences in
LFD amplitudes is done in a rotationally invariant
way~\cite{kms_07}. In nonperturbative LFD calculations, which
always imply Fock space truncation, the dependence on the
light-front plane orientation may survive even in calculated
physical amplitudes. The identification of such amplitudes with
observable quantities becomes therefore ambiguous.

The use of our FSDR scheme in CLFD allows us to answer both
questions. In this scheme, each of the original bare parameters
has an additional index depending on the number of particles in
the Fock sector in which this bare parameter appears. In the
Yukawa model the fermion mass counterterm, $\delta m$, and the
bare coupling constant $g_0$, are thus extended each to a whole
sequence:
\begin{subequations}
\label{fsdr}
\begin{eqnarray}
\label{g0l} g_0 &\to&  g_{0l}\ ,\\
\label{dml} \delta m  &\to& \delta m_l,
\end{eqnarray}
\end{subequations}
with $l=1,2,\ldots N$.
By definition, $g_{01}=0$ and $\delta m_1=0$. For $l\geq 2$,
the quantities $g_{0l}$ and $\delta m_l$
are calculated by solving the systems of equations for the vertex
functions in the $N=2$, $N=3$, ... approximations
successively.

Besides that, as we shall see in the next section, new counterterms which depend explicitly on the
orientation of the light-front plane (i.~e. on $\omega$) should be introduced in order to restore,
if necessary, the rotational invariance
broken by the truncation. In this case,
such counterterms are also mandatory in order to fulfil the
renormalization condition. Similarly to the "traditional"
counterterms, they have Fock sector dependence.
The full set of rules for the calculation of the bare parameters can be found in
Refs.~\cite{kms_08,mstk}.

We emphasize that the FSDR scheme is a general method to make
nonperturbative calculations in truncated Fock space. It can be
easily applied to any physical system admitting particle
counting. The Yukawa model studied in the present work has been
chosen as an illustration of the capabilities of our approach.

%%%%%%%%%%%%%%%%%%%%%%%%%%%%%%%%%%%%%%%%
\section{Renormalization conditions} \label{renor}
Once the bare
coupling constant and the mass counterterms  have been identified,
one should fix them from a set of renormalization conditions. In
perturbation theory, there are three types of quantities to
determine: the mass counterterms and the bare coupling constant just mentioned, and
the field strength renormalization constants. In the on-mass-shell
renormalization scheme, the following conditions are used. The
mass counterterms are fixed from the requirement that the two-point
Green's functions have a pole at the physical masses. The field normalization
constants are fixed from the condition that the residues of the two-point
Green's functions at these poles
equal $1$. The bare coupling constant
is determined by requiring that the on-mass-shell three-point
Green's function is given by the product of the physical
coupling constant and the elementary vertex.

The renormalization conditions in LFD are of slightly different
form, although they rely on the same grounds. The mass counterterm
for each physical state is fixed from the eigenstate
equation~(\ref{kt15}) by demanding that the mass, $M$, of the
physical bound state be identical to the constituent mass $m$. The
bare coupling constant is determined by relating the
on-energy-shell two-body vertex function $\Gamma_2$ to the
physical coupling constant $g$. Finally, the normalization of the
state vector is fixed from the condition ~(\ref{normep}).

In order to set up the relationship between $\Gamma_2$ and the
physical coupling constant, one needs to discuss carefully the
renormalization factors of the external legs of the two-body
vertex function~\cite{hiller,kms_10}. These renormalization factors do
also depend on the order of truncation of the Fock space. In the
Yukawa model, this relationship reads
\begin{equation} \label{gamma2r}
\Gamma_2^{(N)}(s_2=M^2)=g \sqrt{I_{1}^{(N-1)}} \sqrt{Z_b}.
\end{equation}
Eq.~(\ref{gamma2r}) can be interpreted in simple physical terms. Each
leg of the two-body vertex function contributes for a different
factor $\sqrt{Z}$ to the physical coupling constant, where $Z$ is
the field strength normalization factor. The initial fermion state
is the physical state normalized to $1$, so that $Z=1$ in that
case. The final boson line should be renormalized by a factor
$\sqrt{Z_b}$. In the approximation where fermion-antifermion loop
contributions are not considered (the so-called quenched
approximation), we have $Z_b = 1$. Finally, the field strength
normalization factor of the constituent fermion is just the weight
of the one-body component in the norm of the physical
state~\cite{kms_10}, i.e. $Z=I_1$ in that case, according to
Eq.~(\ref{Inor}). Following our FSDR scheme, the normalization
factor of the final constituent fermion should correspond to the
truncation of order $N-1$ of the Fock space, since there is, by
construction of the two-body vertex function, one extra boson in
flight in the final state.

Under the PV regularization, PV particles are considered, in the
interaction Hamiltonian, on equal grounds with the physical ones.
From here it follows that each constituent
particle line in the two-body vertex may correspond to either a
physical or a PV particle. Observable amplitudes are described by
diagrams with physical external legs only. For this reason, the
renormalization condition~(\ref{gamma2r}) should be imposed on the
two-body vertex function with  constituent lines corresponding to
the physical fermion and boson.

The condition~(\ref{gamma2r}) has an important consequence: the
two-body vertex function at $s_2=M^2$ should be independent of the
orientation $\omega$ of the light-front plane. With the spin
decomposition~(\ref{g2}), this implies that the component $b_2$ at
$s_2=M^2$ should be identically zero:
\begin{equation} \label{b2oms}
b_2^{(N)}(s_2=M^2)\equiv 0.
\end{equation}
If Eq.~(\ref{b2oms}) is satisfied, Eq.~(\ref{gamma2r}), in the
quenched approximation, turns into
\begin{equation} \label{b1oms}
b_1^{(N)}(s_2=M^2)\equiv g \sqrt{I^{(N-1)}_{1}}.
\end{equation}

While the property~(\ref{b2oms}) is automatically verified in the
case of the two-body Fock space truncation, provided one uses a
rotationally invariant regularization scheme~\cite{kms_07,kms_08},
this is not guaranteed for higher order truncations. Indeed,
nothing prevents $\Gamma_2$ to be $\omega$-dependent, since it is
an off-shell amplitude, but this dependence must completely
disappear on the energy shell. It would be so if no Fock space
truncation has been done. The latter results in some
$\omega$-dependence of $\Gamma_2$ even on the energy shell. This immediately
makes the general renormalization condition~(\ref{gamma2r})
ambiguous, since its right-hand side is $\omega$-independent.

Another consequence of the truncation of the Fock space is the fact that the components
$b_{1,2}(s_2=M^2)$ are not  constants. Indeed, $b_{1,2}$ depend
{\em a priori} on two kinematical variables. For practical
purposes, we can take the usual longitudinal momentum fraction $x$
and the transverse (with respect to the three-vector ${\bg
\omega}$) momentum ${\bf R}_{\perp}$. They are defined by
\begin{subequations}
\begin{eqnarray} \label{kin}
x&=&\frac{\omega \cd k_2}{\omega \cd p}, \\
{\bf R}_{\perp}&=&{\bf k}_{2\perp}-x{\bf p}_{\perp},
\end{eqnarray}
\end{subequations}
where $k_2$ refers to the momentum of the boson in the two-body
Fock sector. Note that ${\bf R}_{\perp}^2=-(k_2-xp)^2$ is a
relativistic invariant. We have therefore
$b_{1,2}=b_{1,2}(R_\perp,x)$.

We denote by $m$ $(\mu)$ the constituent fermion (boson) mass. The
on-shell condition
\begin{equation}
s_2\equiv\frac{{R}^2_\perp+m^2}{1-x} +\frac{{R}^2_\perp+\mu^2}{x}
=M^2
\end{equation}
can be  used to fix one of the two variables, say $R_\perp$, in
the non-physical domain (for $M=m$):
\begin{equation}\label{R*}
R_\perp = R^*_\perp(x) \equiv i\sqrt{x^2m^2+(1-x)\mu^2},
\end{equation}
so that $b_{1,2}^{(N)}(s_2=M^2)\equiv b_{1,2}^{(N)}(R^*_\perp(x),x)$
calculated in the truncated Fock space depend on
$x$ [an example of a particular form of the function
$b_{2}^{(N)}(R^*_\perp(x),x)$
for $N=3$ is given in Sec.~\ref{anti},
Eq. (\ref{b2inf})], whereas the
conditions~(\ref{b2oms}) and~(\ref{b1oms}) should be valid
identically, i.e. for any value of $x$.

In order to enforce the condition (\ref{b2oms}), we should introduce
 an appropriate counterterm which depends explicitly on the four-vector
$\omega$~\cite{kms_10}. It
originates from the following additional structure in the interaction Hamiltonian:
\begin{equation}
\label{Zomega} \delta {\cal H}^{int}_\omega = -Z_\omega \bar
\psi' \frac{m \sla
\omega}{i \omega \cd
\partial} \psi' \varphi',
\end{equation}
where $Z_\omega$ is just the new counterterm,
$\psi' (\varphi')$ is
the fermion (scalar boson) field, being a sum
of the corresponding physical and PV components, and
$1/(i\omega\cd
\partial)$ is the reversal derivative operator. In the Yukawa
model within the three-body approximation the
contribution~(\ref{Zomega}) is enough to make all renormalization
conditions self-consistent.

In the truncated Fock space, according to the FSDR rules, $Z_{\omega}$
splits into a sequence of Fock sector dependent contributions $Z_{\omega}^{(l)}$,
analogously to the other bare parameters [see Eqs.~(\ref{fsdr})]. For the truncation of order $N$,
it is supposed that all $Z_{\omega}^{(l)}$'s with $l=1,2,\ldots N-1$ have been already known
from lower order truncations, so that we have to determine the "senior" counterterm $Z_{\omega}^{(N)}$
only.
The enforcement of the condition~(\ref{b2oms}),
for any $x$, by an
appropriate choice of the counterterm $Z_\omega^{(N)}$ implies that
the latter should {\em a priori} depend on $x$, i.e. $Z_\omega^{(N)}=Z_\omega^{(N)}(x)$.
If no Fock space truncation occured, we would get
the exact equality $Z_\omega = \mbox{const} \equiv 0$, like e.g. in
perturbation theory.
The same happens for lowest order Fock space truncations because of their triviality. Thus,
$Z_{\omega}^{(1)}=0$ by definition. Then, in the two-body approximation, $Z_{\omega}^{(2)}$ is also zero,
provided the PV regularization is used~\cite{kms_08}. Nonzero and $x$-dependent counterterms $Z_\omega^{(N)}(x)$ appear, starting from $N=3$.

Following the above discussion, the enforcement of the condition~(\ref{b1oms})
induces also a unique dependence of
$g_{0N}=g_{0N}(x)$ as a function of the kinematical variable $x$.

The fact that, in order to satisfy the renormalization conditions,
the bare parameters must depend on the kinematical variable $x$, is
crucial to obtain  results which are finite after the renormalization procedure
in the truncated Fock space is applied. In Sec.~\ref{yukawa}, the stability of
our results relative to the value of the regularization scale,
if the latter reasonably exceeds the physical masses,
will be confirmed numerically with high precision.

At first glance, the $x$-dependence of the bare parameters  seems,
at least, unusual. However, it is a natural consequence of the
truncation. Of course, the bare parameters in the fundamental
non-truncated Hamiltonian are true constants.  After truncation,
the initial Hamiltonian is replaced by a finite matrix which acts
now in a finite Fock space. But it turns out that the modification
of the Hamiltonian is not restricted to a simple truncation.
Indeed, to preserve the renormalization conditions, the
bare parameters in this {\em finite} matrix become $x$-dependent. This
$x$-dependence cannot be derived from the initial Lagrangian. It
appears only after the Fock space truncation.

Our truncated Hamiltonian with $x$-dependent
bare parameters is a self-consistent approximation
to the initial fundamental Hamiltonian. One expects that the approximation becomes better,
when the number of Fock components increases. At the same time,
the $x$-dependence of
the bare parameters should become weaker. We will
see an indication of that behavior in Sec.~\ref{antinum}.
We emphasize that there is no any ambiguity in finding the
bare parameters, in spite of their $x$-dependence. They are completely fixed from the renormalization
conditions.

%%%%%%%%%%%%%%%%%%%%%%%%%%%%%%%%%%%%%%%%%%%%%%%%
%%%%%%%%%%%%%%%%%%%%%%%%%%%%%%%%%%%%%%%%%%%%%%%%

\section{Yukawa model in the $f+fb+fbb$
approximation} \label{ffbfbb}

We apply our general strategy to calculate some physical
observables for the Yukawa model in the truncated Fock space including
sectors with one single fermion, one fermion plus one boson, and
one fermion plus two bosons. Previously, we considered the same
physical system to calculate the fermion AMM~\cite{kms_10}, but without $x$-dependent
bare parameters. The AMM (as well as any calculated observable)
depends on the regularization parameters which are the two PV fermion and boson masses
$m_1$ and $\mu_1$, respectively. In case of a proper renormalization scheme it
must tend to a fixed finite value, when both PV masses become much
greater than the characteristic physical mass scale. In
Ref.~\cite{kms_10} we first took the limit $m_1\to\infty$
analytically, just on the level of the equations for the Fock
components, and then studied the dependence of the AMM on the
remaining PV mass $\mu_1$ numerically. We found, at relatively
small  values of the coupling constant ($\alpha\sim 0.2$), rather good
numerical stability of the AMM as a function of
$\mu_1$, when  $\mu_1\gg m,\,\mu$.

At larger coupling constants ($\alpha\sim 0.5$), we observed weak
but sizable growth of the AMM with the increase of $\mu_1$. A
possible reason of this uncanceled $\mu_1$-dependence of the AMM
is the fact that we used constant, i.~e. $x$-independent,
bare parameters. We shall show in this study that taking into account
$x$-dependence of the
bare parameters in the truncated Fock space allows
to remove completely any dependence of observables on the
regularization parameters, even for rather large values of the coupling
constant. We shall thus calculate not only the AMM,
but both electromagnetic form factors as a function of the
momentum transfer squared.

%%%%%%%%%%%%%%%%%%%%%%%%%%%%%%%%%%%%%%%%%%%%%%%%%%%%%%%%
\subsection{Equations for the Fock components}
\label{eqfock}
\begin{figure}[ht!]
\includegraphics[width=20pc]{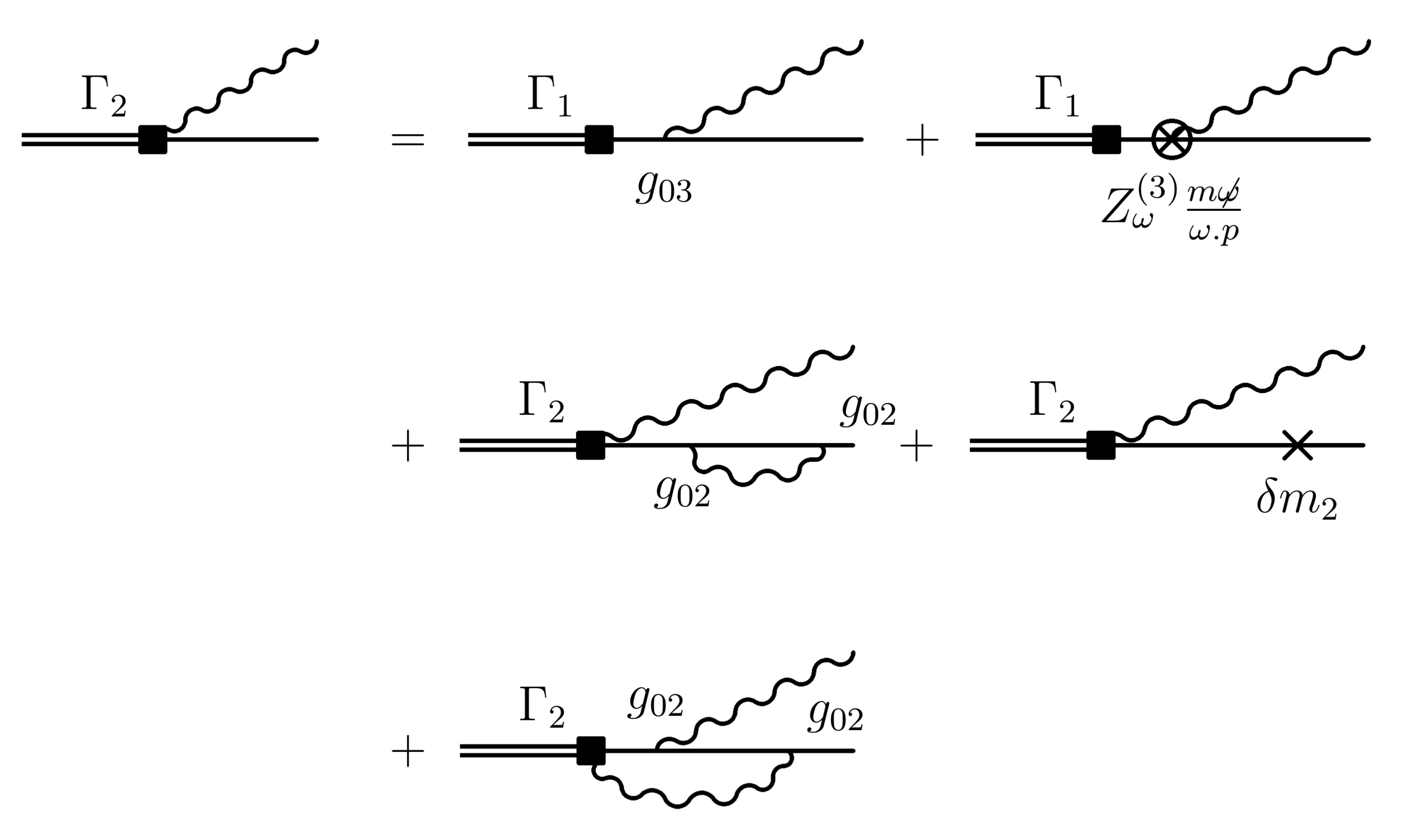}
\caption{Equation for the two-body Fock component in the
$f+fb+fbb$ approximation.} \label{eqgamma2}
\end{figure}

%%%%%%%%%%%%%%%%%%%%%%%%%%%%%%%%%%%%%%%%%%%%%%%%%%%%%%%%
As shown in Ref.~\cite{kms_10}, the system of equations for the
three vertex functions can be reduced to a closed matrix equation
which involves the two-body vertex function $\Gamma_2$ only. This
equation is shown schematically in Fig.~\ref{eqgamma2}. Each
factor at the vertices is taken according to the FSDR scheme
prescriptions. The factors calculated in the three-body
approximation, namely, $g_{03}$ and $Z_{\omega}^{(3)}$ appear in
the amplitudes involving the one-body vertex function $\Gamma_1$
only. Each of the other boson emission and absorbtion vertices in
Fig.~\ref{eqgamma2} brings the factor $g_{02}$, since there exists
one boson in flight at the time moment corresponding to the vertex
and $Z_{\omega}^{(2)}=0$. The mass counterterm contribution
($\delta m_2$) denoted by the cross appears, in the equation for
$\Gamma_2$, within the two-body sector only. Analogous
contributions inside the three-body sector are absent, because
they correspond to the factor $\delta m_1$ which is zero.

The fermion-boson loop (see the third diagram on
the right-hand side of the graphical equation in
Fig.~\ref{eqgamma2}) is nothing else than the two-body fermion
self-energy. In CLFD, it can be represented as the following
general decomposition~\cite{kms_07,kms_10}:
\begin{equation}
\label{selfen2} \Sigma({\sla k})=g_{02}^2\left[ {\cal
A}(k^2)+{\cal B}(k^2)\frac{{\sla k}}{m}+{\cal C}(k^2)\frac{m{\sla
\omega}}{\omega\cd k}\right],
\end{equation}
where ${\cal A}$, ${\cal B}$, and ${\cal C}$ are scalar functions
given in Ref.~\cite{kms_10}. Note that under the PV regularization scheme
we have ${\cal C}(k^2)\equiv 0$.

Since each of the two constituent lines of the two-body vertex may
correspond to either a physical or a PV particle, we have to
distinguish four types of $\Gamma_2$, depending on its "contents":
({\it i}) physical fermion and physical boson;
({\it ii}) physical fermion and PV boson; ({\it iii}) PV fermion
and physical boson; (iv) two PV particles. It is convenient to
supply $\Gamma_2$ by the two indices, $i$ (for fermion) and $j$
(for boson), reflecting the sort of constituent. Each index may
take two values, 0 and 1, corresponding to a physical or a PV
particle, respectively. The substitution $\Gamma_2\to
\Gamma_2^{ij}$ means the substitutions $b_{1,2}\to b_{1,2}^{ij}$
for each of its spin components defined by Eq.~(\ref{g2}). The
particle masses are denoted by $m_i$ and $\mu_j$ with $m_0\equiv
m$ and $\mu_0\equiv \mu$, by definition.

We shall consider in this study two different
orders in taking the infinite limit for the masses of the PV particles.
One of them, already considered in Ref.~\cite{kms_10},
corresponds to the case, when the mass of the PV
fermion, $m_1$, is first set to infinity
(analytically), while the PV boson mass, $\mu_1$,
is kept finite. In the present section, we re-formulate the
system of equations for the Fock components in this limit, taking
into account the $x$-dependence of the bare parameters.
We then study numerically the dependence
of our results on $\mu_1$, when the latter is much greater than
the physical masses. In addition, we
discuss in Sec.~\ref{mu1inf} the opposite order of limits, when $\mu_1$
is first set to infinity.

In order to take the limit $m_1\to\infty$, it is convenient to
introduce   a set of functions, $h_i^j$ and $H_i^j$, which remain
finite in this limit. They are given by~\cite{kms_10}
\begin{equation}
\label{b12hH} b_1^{ij}=\frac{m_i}{m}h_i^j,\,\,\,\,\,\,\,\,
b_2^{ij}=\frac{m_i}{m}\frac{H_i^j-(1-x+\frac{m_i}{m})h_i^j}{2(1-x)}.
\end{equation}
In terms of the functions $h_i^j$ and $H_i^j$,
the equation shown in Fig.~\ref{eqgamma2} corresponds to the
following system of integral equations~\cite{kms_10}\footnote{In
Ref.~\cite{kms_10} we used the functions $\tilde{h}_i^j$ and
$\tilde{H}_i^j$ which differ from $h_i^j$ and $H_i^j$ by their
normalization. The present form of the vertex functions is more
convenient for the determination of $x$-dependent
bare parameters.}:
\begin{subequations}
\label{h0--H1}
\begin{eqnarray}
\label{h0} h_0^j (R_{\perp},x)& = &
g_{03}'+{g'}^2\left[K_1^jh_0^j(R_{\perp},x)+
K_2^jh_1^j(R_{\perp},x)\right]\nonumber\\
&&+{g'}^2 i_0^j(R_{\perp},x),\\
\label{h1} h_1^j (R_{\perp},x)& =
&{g'}^2\left[-K_3^jh_0^j(R_{\perp},x)+
K_4^jh_1^j(R_{\perp},x)\right]\nonumber\\
&&+{g'}^2 i_1^j(R_{\perp},x),\\
\label{H0}
H_0^j (R_{\perp},x)& = &g'_{03}\left[(2-x)+Z'_{\omega}(1-x)\right]\nonumber\\
&&+{g'}^2\left[K_1^jH_0^j(R_{\perp},x)+
K_2^jH_1^j(R_{\perp},x)\right]\nonumber\\
&&+{g'}^2I_0^j(R_{\perp},x) ,\\
\label{H1}
H_1^j (R_{\perp},x)& = & g_{03}'\nonumber\\
&&+{g'}^2\left[-K_3^jH_0^j(R_{\perp},x)+
K_4^jH_1^j(R_{\perp},x)\right]\nonumber\\
&&+{g'}^2I_1^j(R_{\perp},x).
\end{eqnarray}
\end{subequations}
In the above equations we use the following notations. The
coefficients $K_{1-4}^j$ are defined by
\begin{eqnarray}
K_1^j&=&\frac{1}{m}\left\{{\cal B}_r(s_1)-
\frac{2m^2[{\cal A}_r(s_1)+{\cal B}_r(s_1)]}{m^2-s_1}\right\},\nonumber\\
K_2^j&=&\frac{{\cal A}_r(s_1)+{\cal B}_r(s_1)}{m},\nonumber\\
K_3^j&=&\frac{m[{\cal A}_r(s_1)+{\cal B}_r(s_1)]}{m^2-s_1},\nonumber\\
K_4^j&=&\frac{{\cal B}_r(s_1)}{m},\nonumber
\end{eqnarray}
where the subtracted fermion self-energy functions are ${\cal
A}_r(s_1)={\cal A}(s_1)-{\cal A}(m^2)$, ${\cal B}_r(s_1)={\cal
B}(s_1)-{\cal B}(m^2)$, and their argument
\begin{equation}
\label{s1} s_1=-\frac{R_{\perp}^2}{x}+(1-x)m^2
-\frac{1-x}{x}\mu_j^2.
\end{equation}
For convenience, we introduce the quantity
\begin{equation}
\label{gprime} {g'}^2=\frac{g^2}{1+g^2z_0}
\end{equation}
and re-define the bare parameters by
\begin{subequations}
\begin{eqnarray}
\label{eqg03p}
g'_{03}&=&g_{03}\alpha_0\left(\frac{1-g^2\bar{I}_2^{(2)}}{1+g^2z_0}\right),\\
Z'_{\omega}&=&\frac{2Z_{\omega}^{(3)}}{g_{03}}-\frac{\alpha_1}{\alpha_0},\label{eqZomp}
\end{eqnarray}
\end{subequations}
where
\begin{equation}
\label{J2} \bar{I}_2^{(2)}=-\frac{{\cal B}(m^2)}{m}-z_0,
\end{equation}
with
\begin{equation}
\label{z0} z_0=2m\left[{\cal A}'(m^2)+{\cal B}'(m^2)\right].
\end{equation}
The primes at ${\cal A}$ and ${\cal B}$ denote their derivatives
over $s_1$. The quantities $\alpha_0$ and
$\alpha_1$ are, respectively, the physical and PV components of
the one-body (fermion) vertex, introduced in
Ref.~\cite{kms_10}. We do not need to calculate them explicitly in
our study. We can therefore include them into the definitions of
$g'_{03}$ and $Z'_{\omega}$. The quantity $\bar{I}_2^{(2)}$,
multiplied by $g^2$, is nothing else than the norm of the two-body
sector, calculated for the two-body ($f+fb$) Fock space
truncation. Transforming the equation for $\Gamma_2$ to the system
of equations~(\ref{h0--H1}), we took into account the
values of the bare coupling constant $g_{02}$ and the mass
counterterm $\delta m_2$, obtained from the previous
calculations~\cite{kms_08} within the $f+fb$ truncated Fock
space\footnote{In Ref.~\cite{kms_08} the quantity
$\bar{I}_2^{(2)}$ was denoted by $J_2$. We changed the notation
here in order to avoid its confusion with the notation of the
electromagnetic current operator (see below).}:
\begin{subequations}
\begin{eqnarray}
\label{g02} g_{02}^2&=&\frac{g^2}{1-g^2\bar{I}_2^{(2)}},\\
\label{dm2} \delta m_2&=&g_{02}^2\left[{\cal A}(m^2)+{\cal
B}(m^2)\right].
\end{eqnarray}
\end{subequations}
The integral terms describing the contributions of the three-body state to
$\Gamma_2$ are
\begin{subequations}
\label{inIn}
\begin{eqnarray}
i_{n}^j(R_{\perp},x)& =&\int_{0}^{\infty}
 R'_{\perp}dR'_{\perp}\int_0^{1-x} dx' \sum_{i,j'=0}^{1}(-1)^{j'} \nonumber\\
&&\times\left[c_{ni}h_i^{j'}(R'_{\perp},x')+
C_{ni}H_i^{j'}(R'_{\perp},x')\right],\nonumber\\
&&\label{in}\\
I_{n}^j(R_{\perp},x)& =&\int_{0}^{\infty}
 R'_{\perp}dR'_{\perp}\int_0^{1-x} dx' \sum_{i,j'=0}^{1}(-1)^{j'}\nonumber\\
&&\times\left[c'_{ni}h_i^{j'}(R'_{\perp},x')+
C'_{ni}H_i^{j'}(R'_{\perp},x')\right],\nonumber\\
&&\label{In}
\end{eqnarray}
\end{subequations}
with $n=0,\,1$. The coefficients $c$, $C$, $c'$,
and $C'$, which depend on $R_{\perp}$, $R_{\perp}'$, $x$, $x'$,
$j$, and $j'$, are given in Ref.~\cite{kms_10}.

We should now use the renormalization conditions~(\ref{b2oms})
and~(\ref{b1oms}) in order to determine the bare coupling constant
$g'_{03}$ and the counterterm $Z'_{\omega}$. In terms of the
functions $h_i^j$ and $H_i^j$, these conditions write
\begin{subequations}
\label{renh0H0}
\begin{eqnarray}
h_0^0(R_{\perp}^*,x)&=&g\sqrt{1-g^2\bar{I}_2^{(2)}},\label{renh0}\\
H_0^0(R_{\perp}^*,x)&=&g\sqrt{1-g^2\bar{I}_2^{(2)}}(2-x),\label{renH0}
\end{eqnarray}
\end{subequations}
where $R_{\perp}^*$ is defined by Eq.~(\ref{R*}). Setting
$R_{\perp}=R_{\perp}^*$ and $j=0$ in
Eqs.~(\ref{h0}) and~(\ref{H0}),
we demand the relations~(\ref{renh0H0}) to be valid for arbitrary $0<x<1$. As explained
in the previous section, this necessitates to substitute
$g'_{03}\to g'_{03}(x)$ and $Z'_{\omega}\to Z'_{\omega}(x)$. Using
that at $R_{\perp}=R_{\perp}^*$ and $j=0$ we have $s_1=m^2$, and, hence, $K_1^0=z_0$ and
$K_2^0=0$, we get
\begin{subequations}
\label{barex}
\begin{eqnarray}
\label{g03x}
g'_{03}(x)&=&{g'}^2\left[\frac{\sqrt{1-g^2\bar{I}_2^{(2)}}}{g}-i_0^0(R_{\perp}^*,x)\right],
\nonumber\\
&& \\
g'_{03}(x)Z'_{\omega}(x)&=&{g'}^2\left[\frac{(2-x)i_0^0(R_{\perp}^*,x)-
I_0^0(R_{\perp}^*,x)}{1-x}\right]. \label{Zomx} \nonumber\\
\end{eqnarray}
\end{subequations}
Substituting these quantities back into the system of
equations~(\ref{h0--H1}), we obtain
\begin{widetext}
\begin{equation}
\label{hHm1}
\begin{array}{lllllll}
h_0^j (R_{\perp},x)& = &\eta g & + &
{g'}^2\left[K_1^jh_0^j(R_{\perp},x)+
K_2^jh_1^j(R_{\perp},x)\right]&+&{g'}^2\Delta
i_0^j(R_{\perp},x),\\
 h_1^j (R_{\perp},x)& = &&&
{g'}^2\left[-K_3^jh_0^j(R_{\perp},x)+
K_4^jh_1^j(R_{\perp},x)\right]&+&{g'}^2
i_1^j(R_{\perp},x),\\
H_0^j (R_{\perp},x) &= &\eta
g(2-x)&+&{g'}^2\left[K_1^jH_0^j(R_{\perp},x)+
K_2^jH_1^j(R_{\perp},x)\right]&+&{g'}^2\Delta
I_0^j(R_{\perp},x),\\
H_1^j (R_{\perp},x)& = & \eta
g&+&{g'}^2\left[-K_3^jH_0^j(R_{\perp},x)+
K_4^jH_1^j(R_{\perp},x)\right]&+&{g'}^2\Delta I_1^j(R_{\perp},x),
\end{array}
\end{equation}
\end{widetext}
where
\begin{equation}
\label{kappa} \eta=\frac{\sqrt{1-g^2\bar{I}_2^{(2)}}}{1+g^2z_0}
\end{equation}
and
\begin{subequations}
\label{DeltaiI}
\begin{eqnarray}
\label{deltai0}
\Delta  i_0^j(R_{\perp},x)&=&i_0^j(R_{\perp},x)- i_0^0(R_{\perp}^*,x),\\
\label{deltaI0}
\Delta  I_0^j(R_{\perp},x)&=&I_0^j(R_{\perp},x)- I_0^0(R_{\perp}^*,x),\\
\label{deltaI1} \Delta  I_1^j(R_{\perp},x)&=&I_1^j(R_{\perp},x)-
i_0^0(R_{\perp}^*,x).
\end{eqnarray}
\end{subequations}

The functions $h_i^j$ and $H_i^j$, being a solution of the
inhomogeneous system of equations~(\ref{hHm1}), are properly
normalized. Once we have carried out the
renormalization procedure [resulting in
Eqs.~(\ref{barex})], these equations contain only the physical
coupling constant $g$, since $g'$ is expressed via $g$ by
Eq.~(\ref{gprime}). Let us remind that, in
contrast to the results reported in Ref.~\cite{kms_10}, we  do
not fix here any particular value of $x$ at which the
renormalization conditions are considered. We keep the full
$x$-dependence of the bare coupling constant and the
$\omega$-dependent counterterm, according to
Eqs.~(\ref{barex}).

Each index $i$ and $j$ can take two values,
so that we have to deal with eight vertex
functions. The convergence of the integrals over $dR'_{\perp}$ in
the integral terms~(\ref{inIn}) is ensured by the mutual cancellation of the
physical and PV components at $R'_{\perp}\to\infty$ due to the
following properties:
$$
h_i^0(R_{\perp}\to\infty,x)=h_i^1(R_{\perp}\to\infty,x),
$$
$$
H_i^0(R_{\perp}\to\infty,x)=H_i^1(R_{\perp}\to\infty,x),
$$
which automatically follow, when $R_{\perp}\gg \mu_1$, from the
structure of the equations~(\ref{hHm1}). This is a direct consequence of the PV
regularization scheme. However, if one tries to go over to the
limit $\mu_1\to\infty$ inside the integrals (i.~e. before the
integration over $dR'_{\perp}$), some
of them become divergent, unless all the
functions $h_i^j$ and $H_i^j$ vanish, when $R'_{\perp}\to\infty$. But
because of the nonzero free part on the right-hand side of
Eqs.~(\ref{hHm1}) there is no any reason to expect such a behavior
of the solution. For instance, assuming that $h_i^j$ and $H_i^j$
at $R'_{\perp}$ of order $\mu_1$ or greater have some finite
values, we would encounter divergences like $\log \mu_1$. The
renormalization leading to the
subtractions~(\ref{DeltaiI}) also does not fully
prevent from $\mu_1$-divergences, because it "improves" the
asymptotic ($R'_{\perp}\to\infty$) behavior of the integrands only
in the two terms, $\Delta i_0^0(R_{\perp},x)$ and
$\Delta I_0^0(R_{\perp},x)$ (corresponding to the components with
the physical external particles) among the eight ones.  As a
result, we can not take the limit $\mu_1\to\infty$ directly in the
system of equations~(\ref{hHm1}), and the solution essentially
depends on $\mu_1$.

From a practical point of view, we should solve
the equations at large but finite values of $\mu_1$,
express physical amplitudes
(in which, by definition, all the external lines correspond to
physical particles) through the vertex functions, and calculate
observables. We then repeat these steps, gradually increasing
$\mu_1$. If the calculated observables, as a function of $\mu_1$,
tend to stable values within the
required accuracy of
the numerical calculations,
we shall conclude that our renormalization scheme
is successful. Let us emphasize that the stability is
analyzed with respect to observable quantities only, while the
vertex functions, as well as the Fock sector norms may strongly
depend on $\mu_1$, even if it reasonably exceeds the physical
masses. In the following, we will make this procedure
numerically, for the calculation of the electromagnetic form
factors.

%%%%%%%%%%%%%%%%%%%%%%%%%%%%%%%%%%%%%%%%%%%%%%%%%%%%%%%%
\subsection{Calculation of the electromagnetic form factors}
\label{elmfac}
%%%%%%%%%%%%%%%%%%%%%%%%%%%%%%%%%%%%%%%%%%%%%%%%%%%%%%%%
The general decomposition of the spin-1/2 electromagnetic vertex
(EMV) in CLFD is given by~\cite{km96,kms_07,kms_08,kms_10}
\begin{eqnarray}
\bar{u}(p')G^{\rho}u(p)&=&e\bar{u}(p')\left[F_1\gamma^{\rho}+
\frac{iF_2}{2m}\sigma^{\rho\nu}q_{\nu}\right.\nonumber\\
&&+B_1\left(\frac{{\sla \omega}}{\omega\cd
p}P^{\rho}-2\gamma^{\rho}\right)+B_2\frac{m\omega^{\rho}}{\omega\cd
p}\nonumber\\
&&\left.+B_3\frac{m^2{\sla \omega}\omega^{\rho}}{(\omega\cd p)^2}
\right]u(p), \label{EMVF1F2}
\end{eqnarray}
where $P=p+p'$, $q=p'-p$, $\sigma^{\rho\nu}=
i(\gamma^{\rho}\gamma^{\nu}-\gamma^{\nu}\gamma^{\rho})/2$, $e$ is the physical charge,
$F_1$ and $F_2$ are the physical form
factors, and $B_{1,2,3}$ are nonphysical contributions. These
latter originate from possible breaking of rotational
symmetry, caused by the Fock space truncation. Under the condition
$\omega\cd q=0$, all $F_{1,2}$, $B_{1,2,3}$ depend on $Q^2\equiv
-q^2$ only. The physical form factors can be found according
to~\cite{km96}
\begin{subequations}
\label{F1F2em}
\begin{eqnarray}
eF_1&=&\frac{\mbox{Tr}\left[({\sla
p}'+m)\omega_{\rho}G^{\rho}({\sla p}+m){\sla
\omega}\right]}{8(\omega\cd p)^2},
\label{F1}\\
eF_2&=&\frac{m}{2(\omega\cd p)Q^2}
\nonumber\\
&& \times\mbox{Tr}\left[({\sla p}'+m)\omega_{\rho}G^{\rho}({\sla
p}+m)\left(\frac{m{\sla
\omega}}{\omega\cd p}-1\right)\right].\nonumber\\
\label{F2}
\end{eqnarray}
\end{subequations}
Explicit analytical expressions for the two form factors $F_1$ and
$F_2$ in the Yukawa model for the case of the $f+fb+fbb$ Fock
space truncation
are given in Ref.~\cite{kms_10}. For this
reason, we will not dwell on technical derivations, but focus on
how the FSDR scheme works to renormalize the fermion
EMV.

From Eqs.~(\ref{F1F2em}) we can see that both physical
form factors are determined by the contraction of the EMV with the
four-vector $\omega$, sandwiched between the bispinors.
It is thus convenient to define the operator
\begin{equation}
\label{Jph} eJ(Q)=\frac{\bar{u}(p')\,\omega_{\rho}
G^{\rho}\,u(p)}{2(\omega\cd p)}.
\end{equation}
In standard LFD it is nothing else
than the plus-component of the electromagnetic current. Within the
FSDR scheme in truncated Fock space,
the operator~(\ref{Jph}), similarly to
the vertex functions, should be supplied with a
superscript indicating the order of truncation $N$. $J^{(N)}(Q)$ can be represented as
a superposition of contributions $J^{(N)}_n(Q)$ from Fock sectors with
different numbers $n$ of particles:
\begin{equation}
 eJ^{(N)}(Q)=\sum_{n=1}^Ne_{0(N-n+1)}J^{(N)}_n(Q),\label{EMVs}
\end{equation}
where $e_{0(N-n+1)}$ is the electromagnetic
bare coupling constant in the $n$-body sector~\cite{kms_10}.
By definition, $e_{01}=e$. The standard renormalization condition for the EMV
\begin{equation}
 \left.G^{\rho}\right|_{p'=p}=e\gamma^{\rho}\label{EMVrenc}
\end{equation}
written in terms of $J(Q)$ has a very simple form
\begin{equation}
 J(0)=1.\label{Jrenc}
\end{equation}
If the norms $I_n$ of all Fock sectors are finite, the following
relation must be valid:
\begin{equation}
 J^{(N)}_n(0)=I^{(N)}_n\label{EMVn}
\end{equation}
for any $N$.
The normalization condition~(\ref{Inor}) just guarantees the
property~(\ref{Jrenc}). In the Yukawa model however, the norm of
each Fock sector is infinite, and the validity of Eq.~(\ref{EMVn})
depends on the regularization scheme. In Ref.~\cite{kms_08} it was
proved that Eq.~(\ref{EMVn}) held true for any
Fock sector containing one fermion and arbitrary number of bosons,
provided the PV regularization is used, when  $m_1\to\infty$ at
finite $\mu_1$. Since this is just the case we consider in
the present section, we can safely use the relation~(\ref{EMVn}). A direct
consequence of the latter is the following result~\cite{kms_08}:
\begin{equation}
 e_{0n}=e,\label{e0n}
\end{equation}
for any $n$, i.~e. the electromagnetic coupling constant is not
renormalized at all.

\begin{figure}[ht!]
\includegraphics[width=8pc]{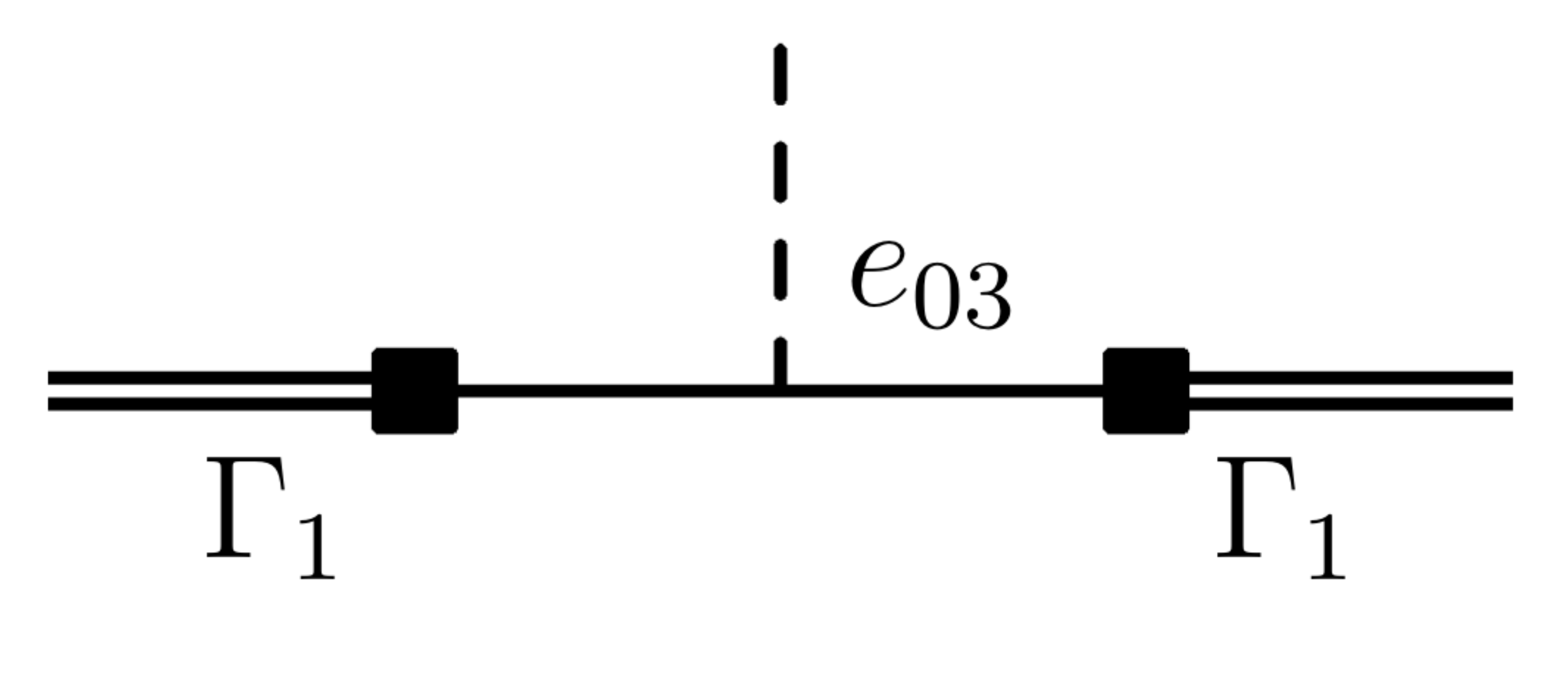}
\caption{One-body ($f$) Fock sector contribution to the
electromagnetic vertex.} \label{ff1}
\end{figure}

\begin{figure}[ht!]
\includegraphics[width=8pc]{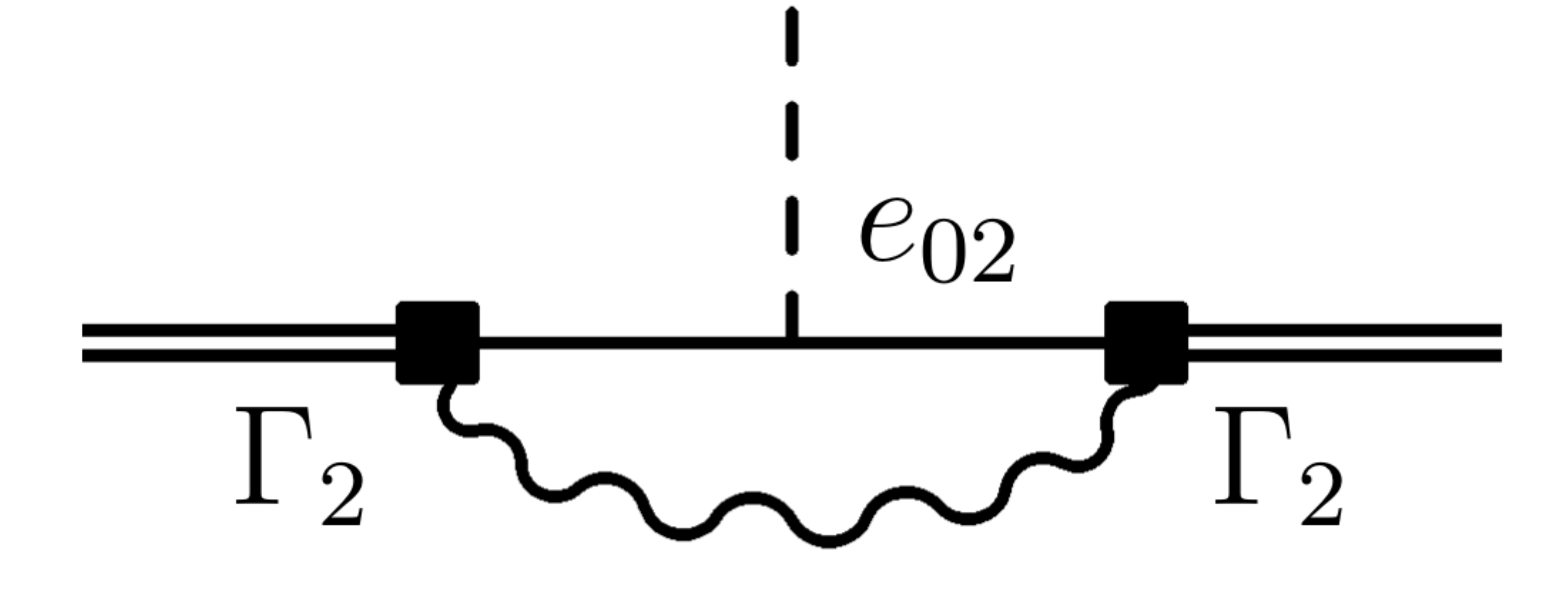}
\caption{Two-body ($fb$) Fock sector contribution to the
electromagnetic vertex.} \label{ff2}
\end{figure}

\begin{figure}[ht!]
\includegraphics[width=20pc]{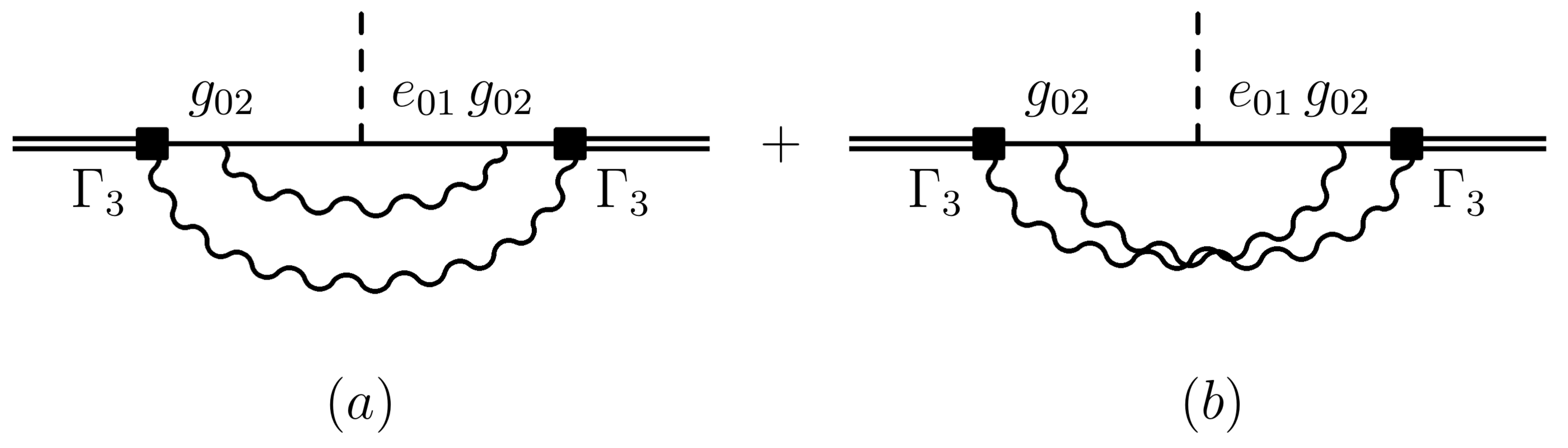}
\caption{Three-body ($fbb$) Fock sector contributions to the
electromagnetic vertex, expressed through the two-body Fock
component, with a nested fermion-boson loop (a)
and with crossed boson loops (b).} \label{ff3}
\end{figure}

In the $f+fb+fbb$ approximation, the EMV is given by
a sum of contributions shown graphically in
Figs.~\ref{ff1}--\ref{ff3}. The diagrams in Figs.~\ref{ff1}
and~\ref{ff2} correspond, respectively, to the one- and two-body
sector contributions to the EMV. The three-body sector
contribution in Fig.~\ref{ff3} is expressed, for convenience,
through the two-body vertex and separated into two parts (a) and
(b). We will denote these particular contributions to the full
operator $J(Q)$ as $J_1(Q)$, $J_2(Q)$, $J_{3a}(Q)$, and
$J_{3b}(Q)$. To shorten notations, hereafter we omit
the superscript $N$ for all quantities related to the three-body
approximation, keeping it however, if a given quantity is calculated
within a truncation of another order. From Eq.~(\ref{EMVs}) we find
\begin{equation}
 eJ(Q)=e_{03}I_1+e_{02}J_2(Q)
+e\left[J_{3a}(Q)+J_{3b}(Q)\right].\label{EMV1}
\end{equation}
Due to Eqs.~(\ref{Jrenc}) and~(\ref{e0n}), the renormalized $J(Q)$ becomes
\begin{eqnarray}
 J(Q)&=&1+\left[J_2(Q)-J_2(0)\right]
\nonumber\\
&&+\left[J_{3a}(Q)+J_{3b}(Q)-J_{3a}(0)-
J_{3b}(0)\right].\label{EMVren}
\nonumber\\
\end{eqnarray}
It is convenient to separate the common
normalization factor
$\sqrt{1-g^2\bar{I}_2^{(2)}}$
($\bar{I}_2^{(2)}$ logarithmically diverges,
when $\mu_1\to\infty$) from the vertex functions
by defining a new set of functions
\begin{equation}
\label{hHbar} \left(
\begin{array}{c}
\bar{h}_{0,1}^j\\
\bar{H}_{0,1}^j
\end{array}
\right)= \frac{1}{g\sqrt{1-g^2\bar{I}_2^{(2)}}}
\left(
\begin{array}{c}
h_{0,1}^j\\
H_{0,1}^j
\end{array}
\right).
\end{equation}
We will supply with a bar each contribution to $J(Q)$, calculated
through the functions $\bar{h}_i^j$ and $\bar{H}_i^j$ in the
vertices. In $J_{3a}(Q)$ and $J_{3b}(Q)$ we suppose that the
factor $g_{02}^2$ coming from the internal vertices in
Fig.~\ref{ff3} is also separated. Eq.~(\ref{EMVren}) transforms thus
to
\begin{eqnarray}
 J(Q)&=&1+g^2(1-g^2\bar{I}_2^{(2)})\left[\bar{J}_2(Q)-
\bar{J}_2(0)\right]\nonumber\\
&&+g_{02}^2\ g^2\ (1-g^2\bar{I}_2^{(2)})\\
&&\times\left[\bar{J}_{3a}(Q)+\bar{J}_{3b}(Q)-\bar{J}_{3a}(0)-
\bar{J}_{3b}(0)\right]\nonumber.
\label{EMVg}
\end{eqnarray}
Substituting here $g_{02}^2$ from Eq.~(\ref{g02}) and rearranging
the order of terms, we obtain
\begin{eqnarray}
 J(Q)&=&1+g^2\left[\bar{J}_2(Q)-
\bar{J}_2(0)\right]\nonumber\\
&&+g^4\left\{\left[\bar{J}_{3a}(Q)-\bar{I}_2^{(2)}\bar{J}_2(Q)\right]\right.
\nonumber\\
&&-\left.\left[\bar{J}_{3a}(0)-\bar{I}_2^{(2)}\bar{J}_2(0)\right]\right\}
\nonumber\\
&&+g^4\left[\bar{J}_{3b}(Q)-\bar{J}_{3b}(0)\right]. \label{EMVgf}
\end{eqnarray}
Let us emphasize that although Eq.~(\ref{EMVgf}) looks like a
perturbative expansion, it has no relation to perturbation theory,
since the quantities $\bar{J}_2$, $\bar{J}_{3a}$, and
$\bar{J}_{3b}$ have rather complicated dependence on the coupling
constant $g$, governed by the nonperturbative equations for the
vertex functions. Nevertheless, it is rather instructive to study
how $\mu_1$-divergences disappear in Eq.~(\ref{EMVgf}), when the
vertex functions are given by their perturbative values:
$$
\bar{h}_0^j=\bar{H}_1^j=1,\quad\bar{h}_1^j=0,\quad\bar{H}_0^j=2-x.
$$
In this case, the $\mu_1$-dependence of $\bar{J}_2$,
$\bar{J}_{3a}$, and $\bar{J}_{3b}$ comes only from the integrals
for the EMV. It is easy to show that both
$\bar{J}_2$ and $\bar{J}_{3b}$ diverge, when $\mu_1\to\infty$, as $\log\mu_1$ with a
coefficient independent of $Q$. Subtracting from each of them
its value at $Q=0$, we just cancel divergent logarithmic terms,
and the result is finite. Concerning the contribution
$\bar{J}_{3a}$ [see Fig.~\ref{ff3}(a)], it has a nested
fermion-boson loop which diverges, after the
integration over the corresponding kinematical
variables, as $\log\mu_1$. The integration over the variables of
the external loop gives one more $\log\mu_1$, so that
$\bar{J}_{3a}$ diverges like $\log^2\mu_1$. Its renormalization
occurs in two steps, as is seen from the expression in braces in
Eq.~(\ref{EMVgf}). In a first step, we form the difference
$\left[\bar{J}_{3a}(Q)-\bar{I}_2^{(2)}\bar{J}_2(Q)\right]$ which
kills $\log\mu_1$ coming from the nested loop. Indeed, by
definition, $\bar{I}_2^{(2)}$ coincides, up to a factor of $g^2$,
with the two-body normalization integral which, according to
Eq.~(\ref{EMVn}), is related to the two-body contribution to the
EMV by
\begin{equation}
\label{Jb} \bar{J}^{(2)}_2(0)=\bar{I}_2^{(2)}.
\end{equation}
Due to the fact that the divergent part of the nested loop
coincides with that in $\bar{J}^{(2)}_2(0)$, it is completely
canceled in the difference
$\left[\bar{J}_{3a}(Q)-\bar{I}_2^{(2)}\bar{J}_2(Q)\right]$. In a
second step, subtracting from this difference its value at $Q=0$,
we remove divergences coming from the integration over the
external loop variables. In standard perturbation theory, the
renormalization is made by the same scenario. But we emphasize
that Eq.~(\ref{EMVgf}) naturally appears within the FSDR scheme
which is fully nonperturbative.

When the vertex functions are solutions of the system of
equations~(\ref{hHm1}) and depend non-trivially on particle
momenta and on the coupling constant, it is much more difficult to
trace analytically the cancellation of $\mu_1$-divergences in calculated
observables. For this reason, we will make the corresponding analysis numerically.

%%%%%%%%%%%%%%%%%%%%%%%%%%%%%%%%%%%%%%%%%%%%%%%%%%%%%%%%%%%
\subsection{Numerical results} \label{yukawa}
%%%%%%%%%%%%%%%%%%%%%%%%%%%%%%%%%%%%%%%%%%%%%%%%%%%%%%%%%%%
We solve the system of integral
equations~(\ref{hHm1}), in order to calculate the two-body Fock component.
Knowing the latter, we are able to find the three-body component
as well (see Ref.~\cite{kms_10} for details). After that, we
calculate the electromagnetic form factors by means of
Eqs.~(\ref{F1F2em}), with the EMV $G^{\rho}$ given by a
sum of contributions shown in Figs.~\ref{ff1}--\ref{ff3}. The EMV
is expressed through the quantity $J(Q)$ for which we take its
renormalized value~(\ref{EMVgf}). In all computations, we use the
physical particle masses $m=0.938$ and $\mu=0.138$ reflecting the
characteristic nuclear physics mass scales. Each physical quantity
is calculated for  three values of the physical coupling
constant, $\alpha=g^2/4\pi=0.5$, $0.8$, and $1.0$.

\begin{figure}[bth]
\includegraphics[width=8.5cm]{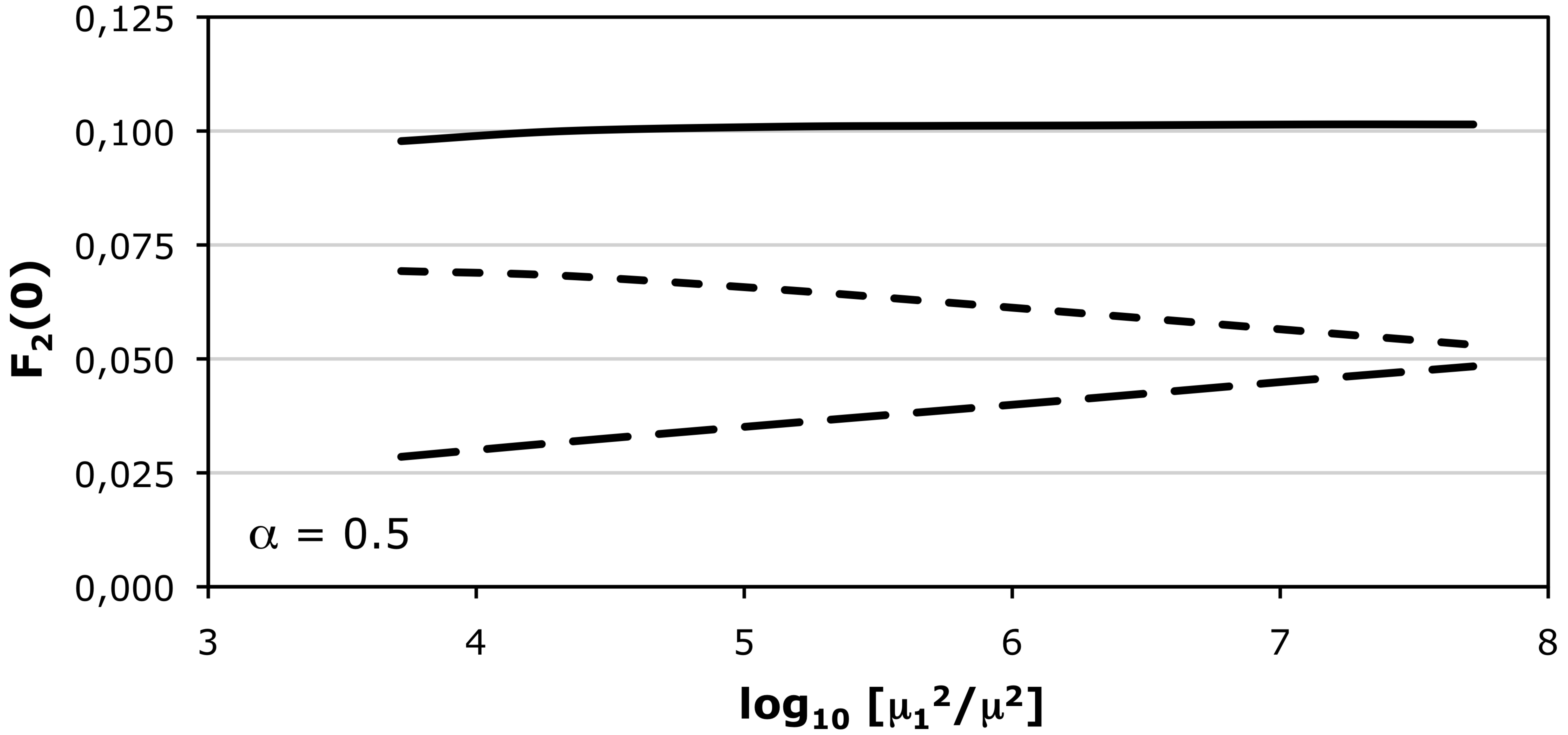}
\includegraphics[width=8.5cm]{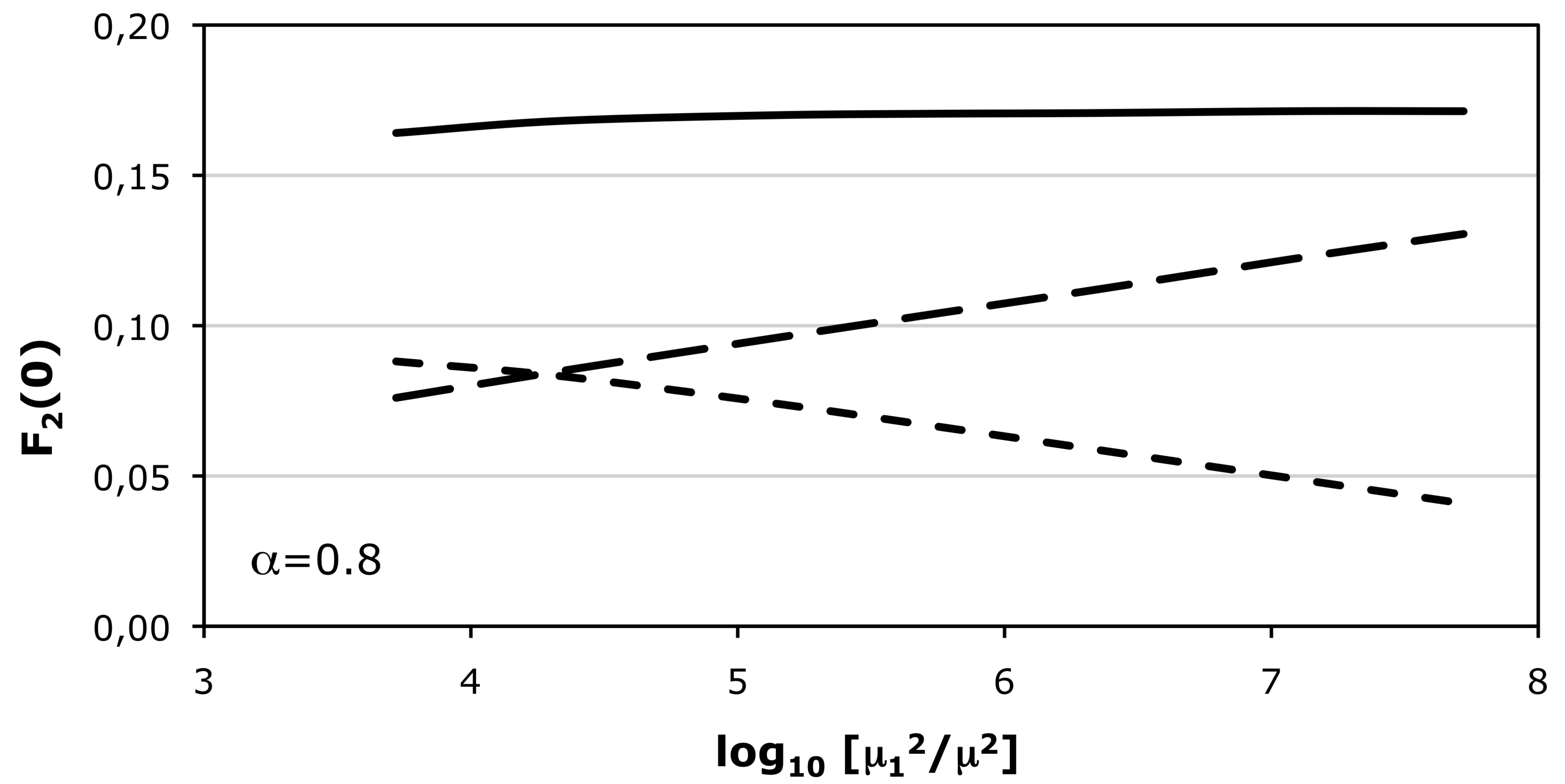}
\includegraphics[width=8.5cm]{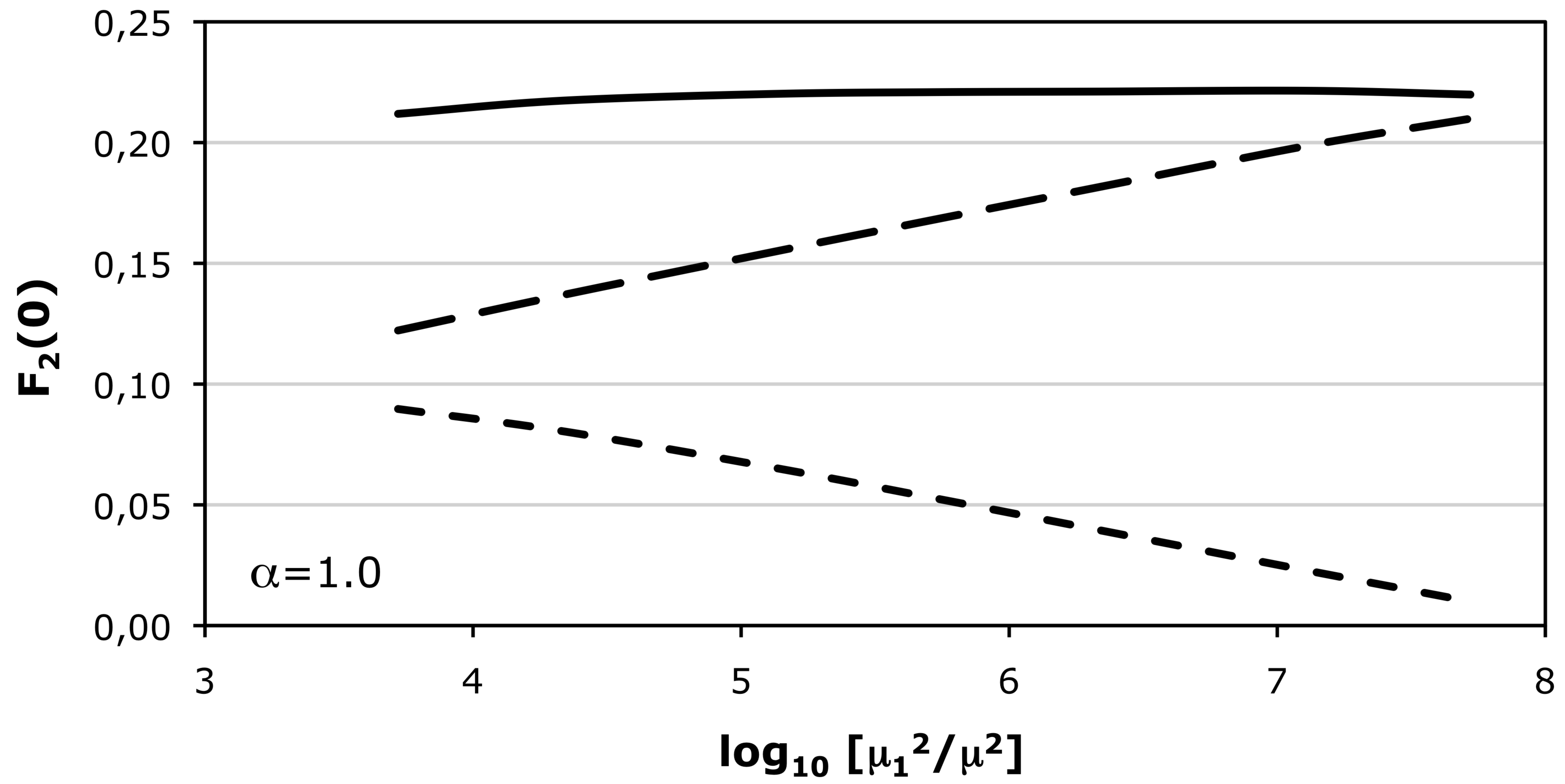}
\caption{The anomalous magnetic moment in the Yukawa model as a
function of the PV mass $\mu_1$, for three
different values of the coupling constant, $\alpha = 0.5$ (upper
plot), $0.8$ (middle plot) and $\alpha = 1.0$ (lower plot). The
dashed and long-dashed lines are, respectively, the two- and three-body
contributions, while the solid line is the total
result.}\label{amm}
\end{figure}

We first compute the fermion AMM which is the
value $F_2(Q^2=0)$. It is shown in  Fig.~\ref{amm} as a function of the PV boson mass $\mu_1$.
Each of the
two- and three-body Fock sector contributions to the AMM
essentially depends on $\mu_1$, while their sum is stable, as
$\mu_1$ becomes large enough. Note that using $x$-dependent bare
parameters removes $\mu_1$-dependence of the AMM, observed in
Ref.~\cite{kms_10} already for $\alpha\sim 0.5$, even for larger coupling constants.

\begin{figure}[bth]
\includegraphics[width=8.5cm]{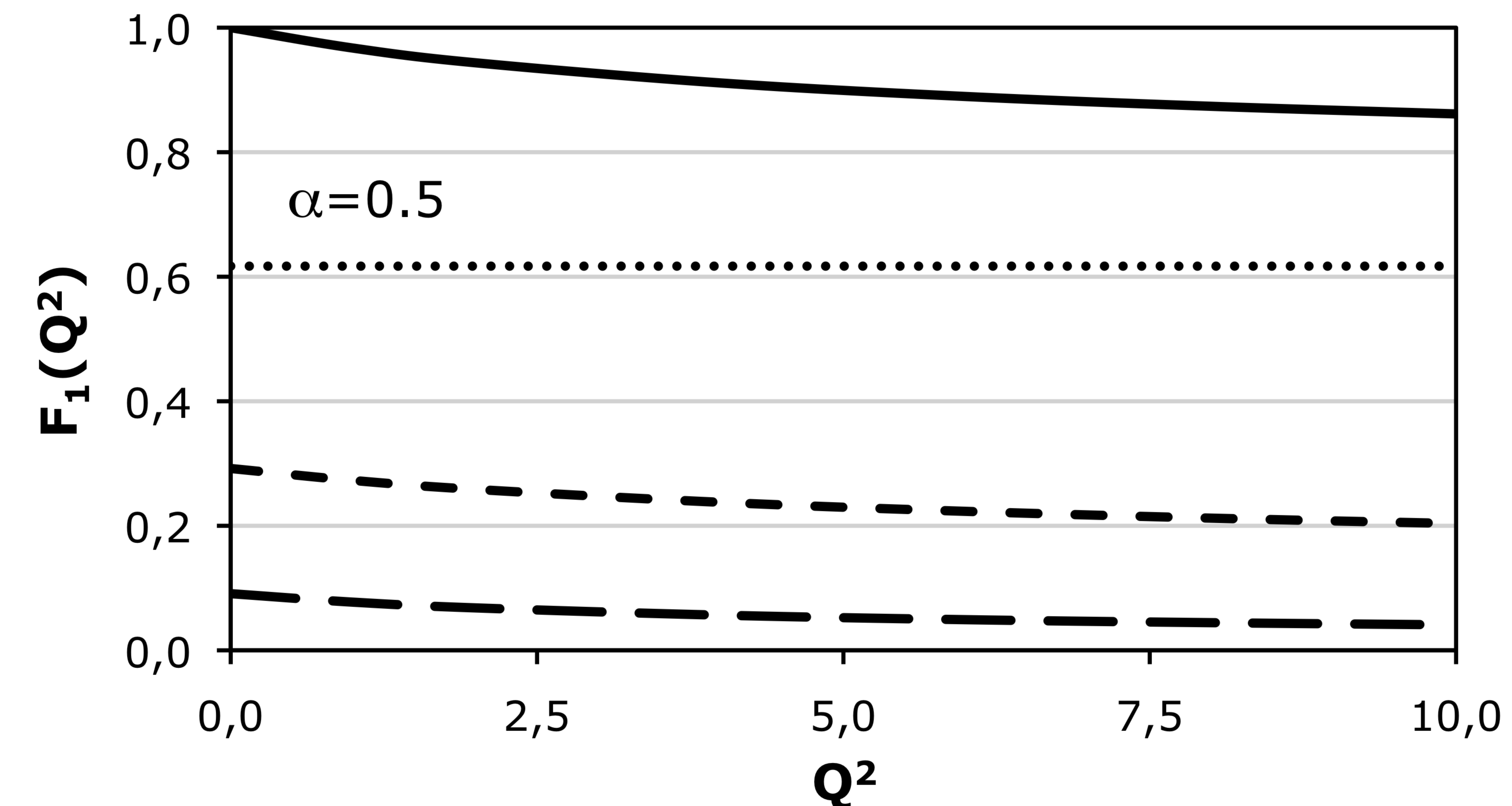}
\includegraphics[width=8.5cm]{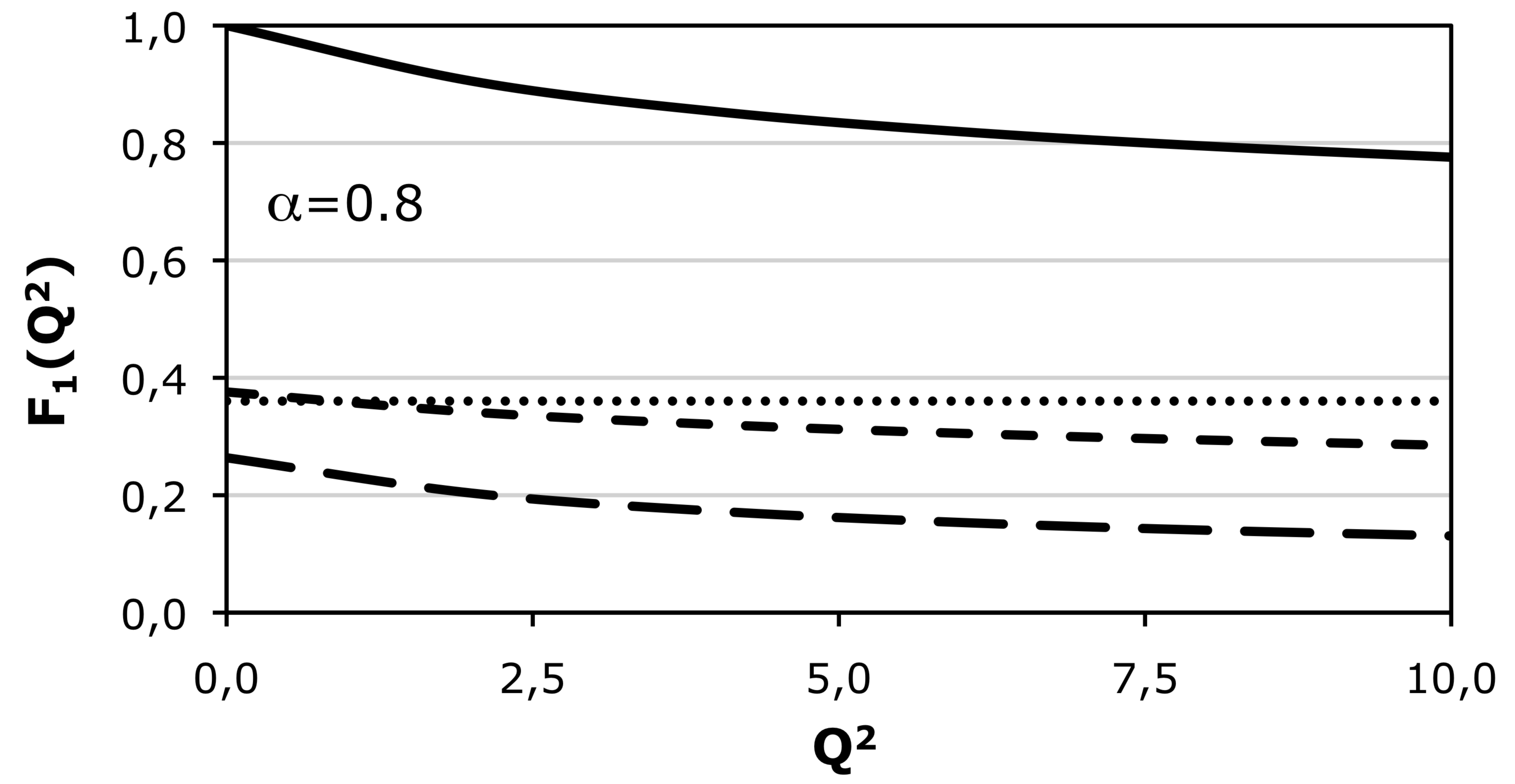}
\includegraphics[width=8.5cm]{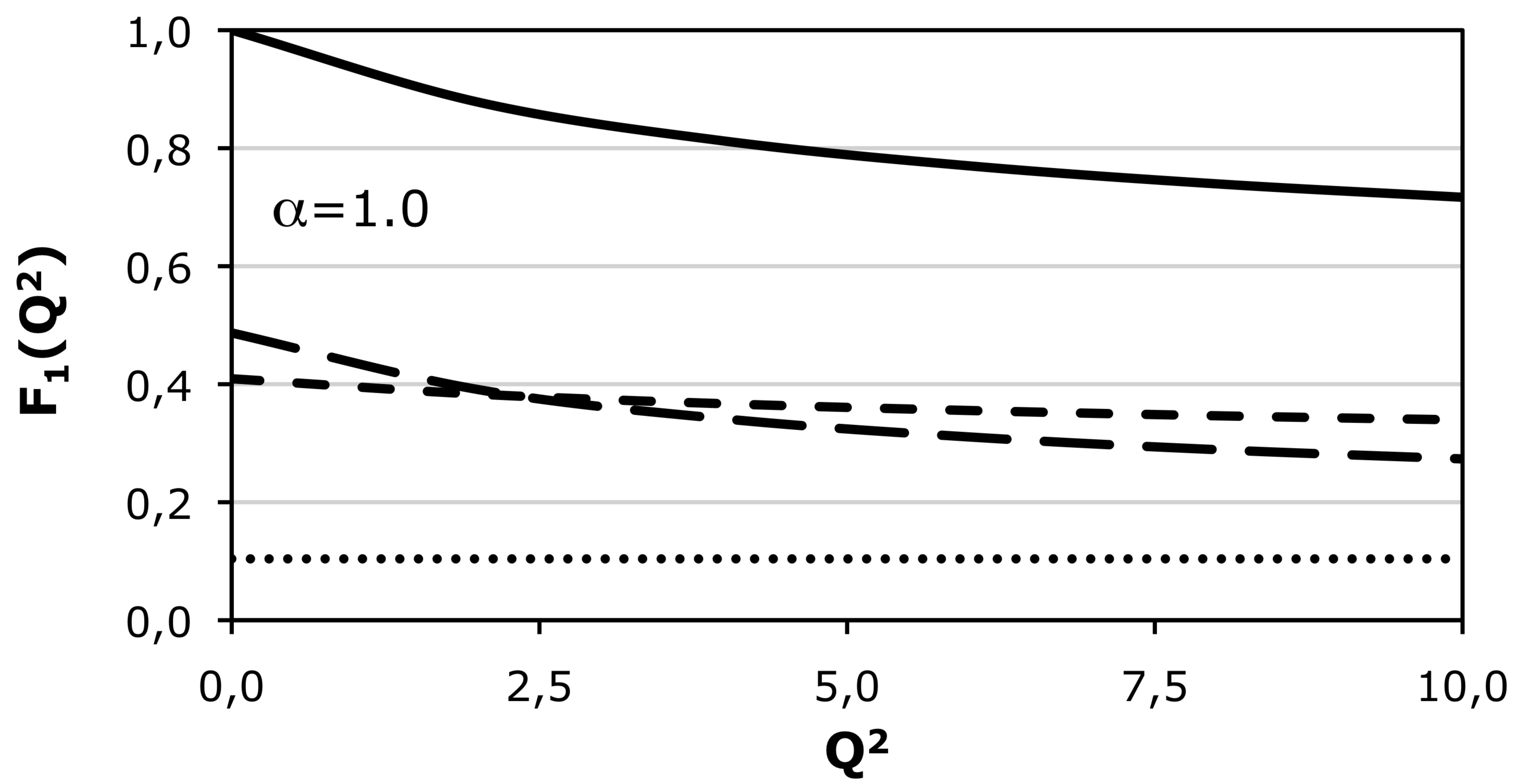}
\caption{Electromagnetic form factor $F_1(Q^2)$ in the Yukawa
model, at $\mu_1=100$, for $\alpha = 0.5$ (upper plot), $0.8$
(middle plot) and $\alpha = 1.0$ (lower plot). The
dotted, dashed, and long-dashed lines are, respectively, the one-, two-, and three-body
contributions, while the solid line is the total
result.}\label{F1fig}
\end{figure}

\begin{figure}[bth]
\includegraphics[width=8.5cm]{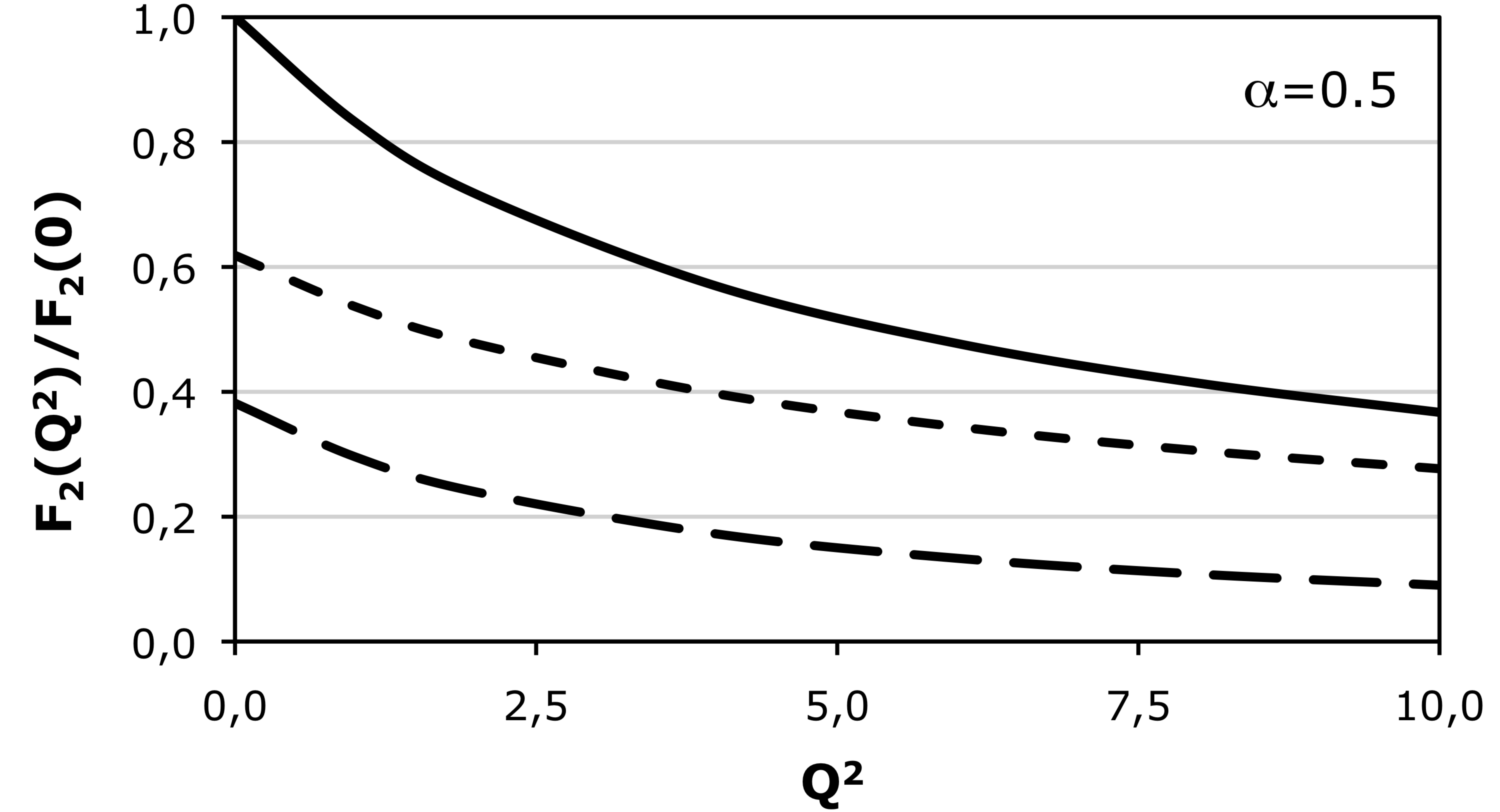}
\includegraphics[width=8.5cm]{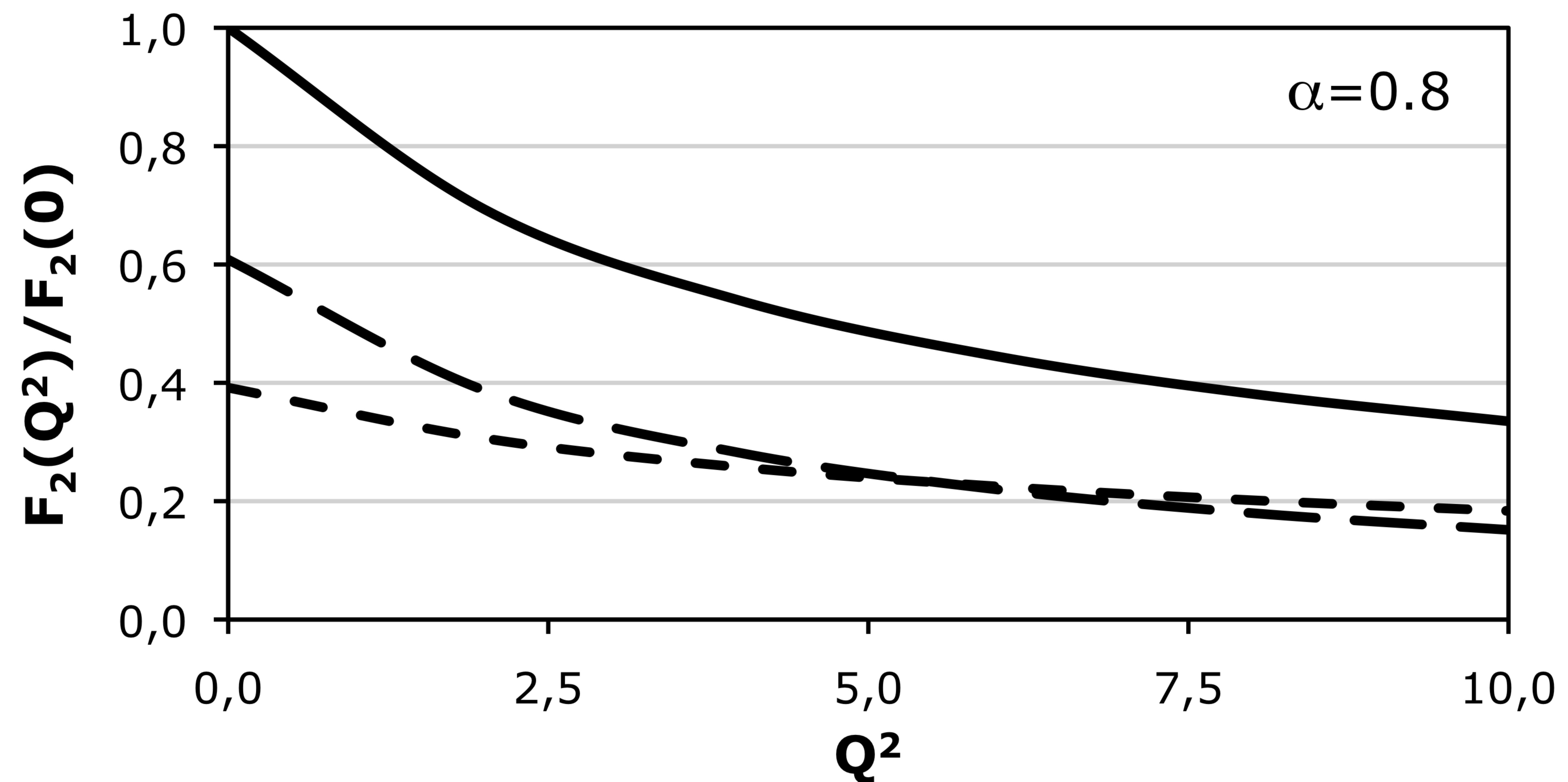}
\includegraphics[width=8.5cm]{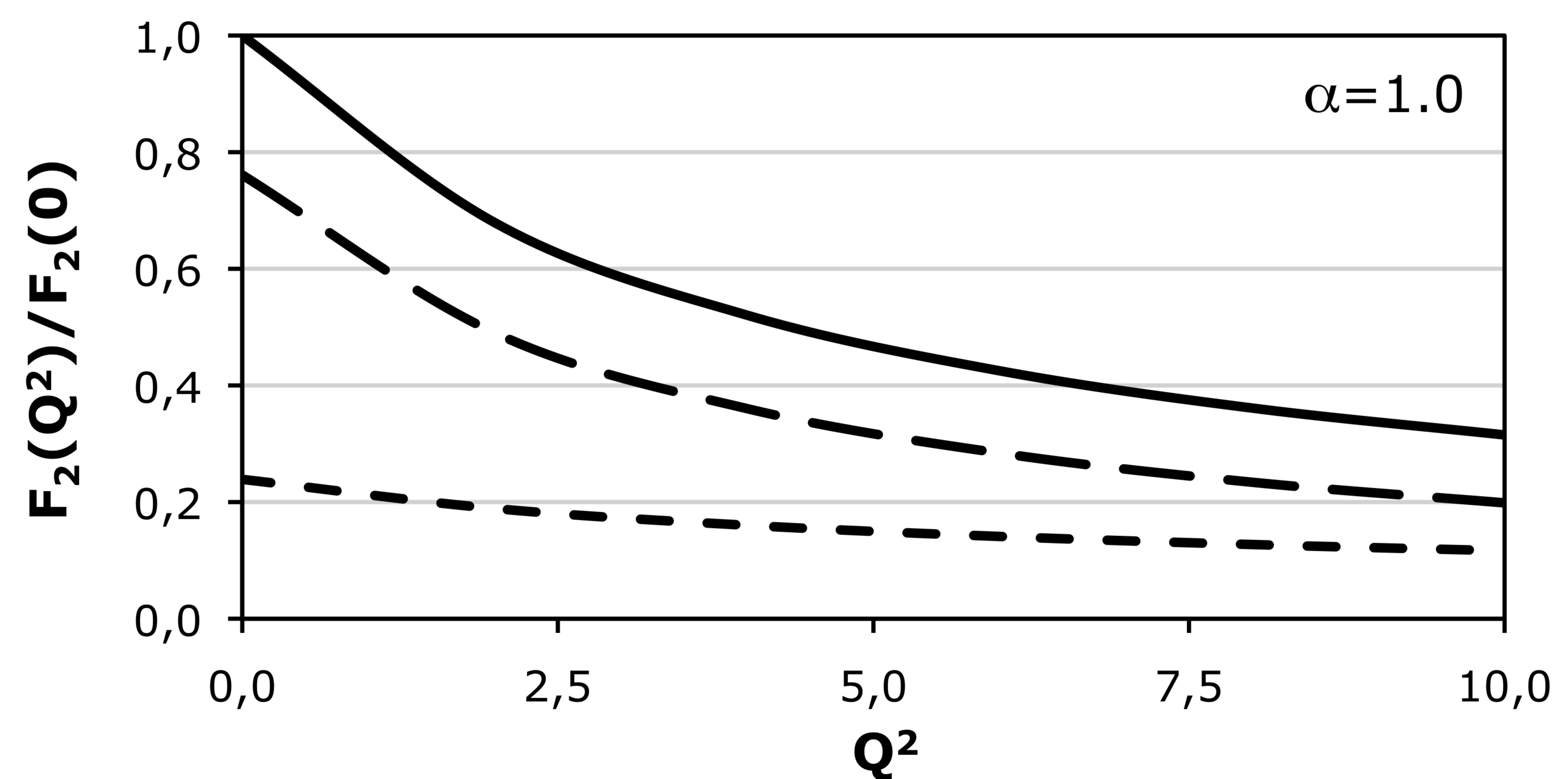}
\caption{The same as in Fig.~\ref{F1fig}, but for the
ratio $F_2(Q^2)/F_2(0)$.}\label{F2fig}
\end{figure}

In Figs.~\ref{F1fig} and~\ref{F2fig}, we plot
the electromagnetic form factors $F_1$ and $F_2$, respectively, as
a function of $Q^2$ for $\mu_1=100$, for the three values of the
coupling constant, considered here. For this value of $\mu_1$, the
form factors already reach the zone of stability: further increase of
$\mu_1$ does not lead to changes distinguishable in the scale of
the figures.


\begin{figure}[bth]
\includegraphics[width=8.5cm]{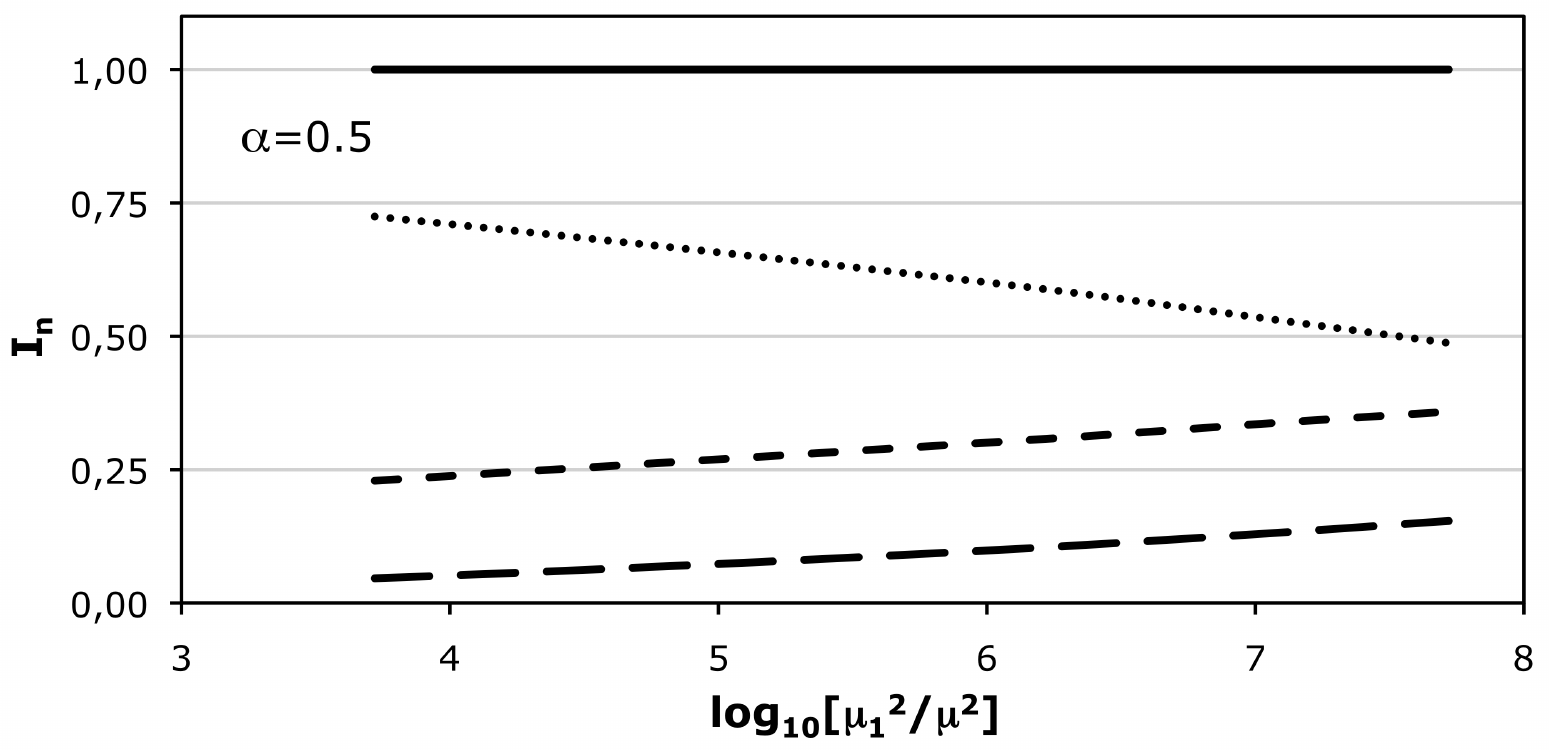}
\includegraphics[width=8.5cm]{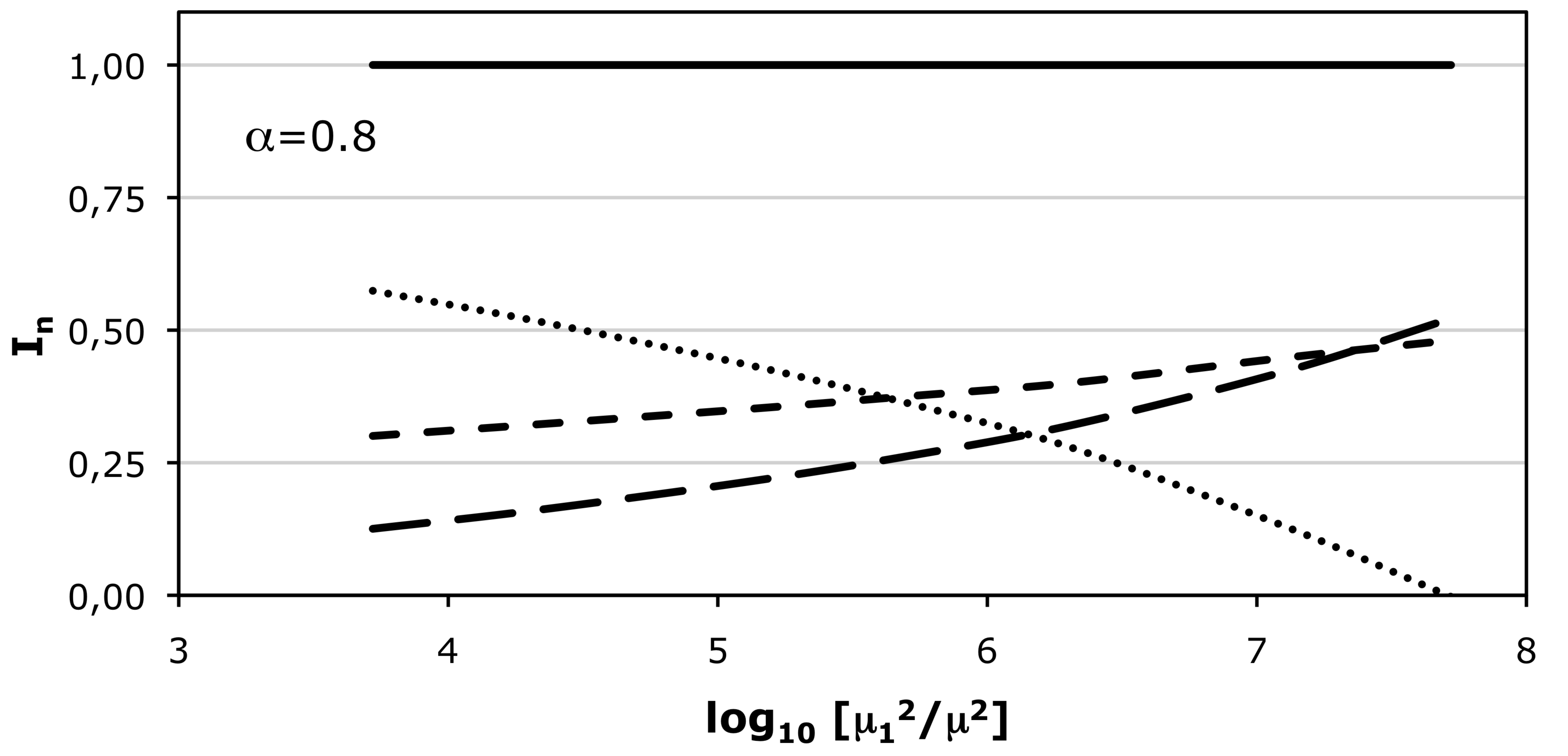}
\includegraphics[width=8.5cm]{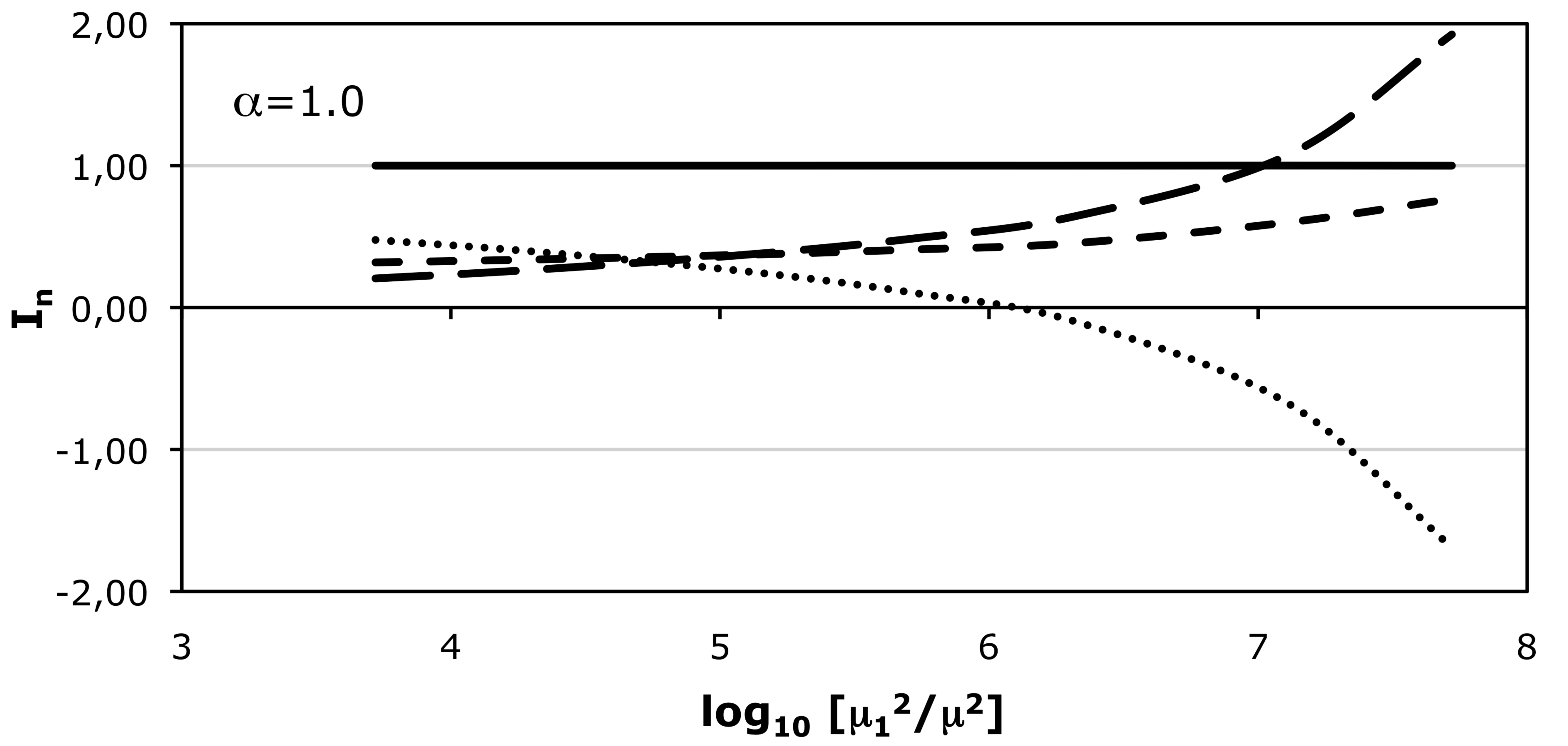}
\caption{Individual contributions of the one- (dotted line),
two- (dashed line), and three-body (long-dashed line) Fock sectors to
the norm (solid line) of the state vector, as a
function of the PV boson mass $\mu_1$, for $\alpha = 0.5$ (upper
plot), $0.8$ (middle plot), and $\alpha = 1.0$ (lower
plot).}\label{n123}
\end{figure}

We finally show in Fig.~\ref{n123} the contributions of the one-,
two-, and three-body sectors to the norm of the state vector. It
is interesting to note that for large enough values of the PV
boson mass $\mu_1$, the norm of the one-body Fock sector gets
negative, while the norm of the two-body Fock sector becomes
larger than $1$, the sum of all contributions being fixed to $1$
by construction. However, physical observables are well defined
and do not show any discontinuity, as a function
of $\mu_1$, when $I_1$ changes the sign or $I_2$ exceeds unity.
This is an illustration of the fact that the norms of the Fock components,
in the presence of the PV sectors having  negative
norms, are not physical observables and, hence, they are expected
to be regularization scale dependent. As we can see in these
figures, they do depend on $\mu_1$, unlike the electromagnetic
form factors.
The stability of the form factors
to variations of the
PV masses, when the latter ones are large
compared to the physical masses, opens
encouraging perspectives for our method of nonperturbative
renormalization.

For obvious reasons  related to the finite accuracy of numerical
calculations, this domain of stability can not be checked
up to infinite values of the PV mass $\mu_1$. However,
we have found that for enough large $\mu_1$ the stability
disappears. Namely, when $\mu_1$ approaches some
critical mass $\mu_{1c}$, all the functions $h_i^j$ and $H_i^j$ defining
the two-body Fock component become very large,
when $R_{\perp}\to\infty$ or $x\to 0$. At $\mu_1=\mu_{1c}$
these functions are unbounded, because their asymptotic
values turn into infinity. Since $h_i^j$, $H_i^j$ start growing from the
characteristic values $R_\perp \simeq \mu_1$ and $x
\simeq (m/\mu_1)^2$, the calculated observables
are sensitive to the variations of $\mu_1$, if it is not far enough from $\mu_{1c}$. The critical mass $\mu_{1c}$
which can be calculated analytically
has very sharp dependence on the coupling constant $\alpha$. For
$\alpha\to 0$ it goes to infinity.
As $\alpha$ increases, $\mu_{1c}$ falls down very rapidly.
Thus, for the physical masses we consider ($m=0.938$, $\mu=0.138$)
the values of $\mu_{1c}$, corresponding to $\alpha=0.5$, $0.8$, and $1$,
are, respectively, $5.17\cd 10^8$, $4.17\cd 10^4$, and $1.8\cd 10^3$.
For little bit larger $\alpha$'s, $\mu_{1c}$ becomes comparable
with the physical masses. In this domain, a reliable numerical
calculation of physical observables is impossible. So, in the
Yukawa model, within the three-body truncation and using the PV
regularization scheme,  the typical dependence of calculated
observables on $\mu_1$ looks as follows. At $\mu_1$ comparable
with the physical masses, observables are sensitive to the value
of $\mu_1$. Then, in the region $m\ll\mu_1\ll\mu_{1c}$ this
dependence has a plateau corresponding to the zone of stability.
This is just our "working" region, inside which we can trust  the
physical meaning of our numerical results. At $\mu_1$ close to
$\mu_{1c}$ we can not expect to perform reliable numerical
calculation of physical observables.
The determination of the critical mass $\mu_{1c}$, as well
as studying the properties of the system of
equations~(\ref{hHm1}), when $\mu_1$ is close to $\mu_{1c}$,
go beyond the framework of the present article and will be
explained in detail in a future publication.\\

%%%%%%%%%%%%%%%%%%%%%%%%%%%%%%%%%%%%%%%%%%%%%%%%%%%%%%%%%%%
\subsection{Additional test of the
renormalization scheme}\label{mu1inf}
%%%%%%%%%%%%%%%%%%%%%%%%%%%%%%%%%%%%%%%%%%%%%%%%%%%%%%%%%%%

The system of equations~(\ref{hHm1}) was obtained
in the limit $m_1\to\infty$, while $\mu_1$ being fixed. We then
increase $\mu_1$ till the stability of
calculated observables is
reached. Physical observables however must be independent
of the
order in which the infinite PV mass limit is taken. For this
reason, we shall make a test of the self-consistency of our
renormalization procedure. Namely, we perform an additional
calculation of the AMM, taking first the limit $\mu_1\to\infty$
(analytically) and then the limit $m_1\to\infty$ (numerically).

The corresponding system of equations for the vertex functions in
the limit $\mu_1\to\infty$  can be obtained, analogously to Eqs.~(\ref{hHm1}), from the
general equation shown graphically in Fig.~\ref{eqgamma2}.
Omitting here all
technical details, we
indicate only the main differences of the system of equations
obtained in the limit $\mu_1\to\infty$ from Eqs.~(\ref{hHm1}).

When $\mu_1\to\infty$, the four vertex functions with $j=0$
form a closed sub-system of linear
integral equations, while the other four equations involving the
functions with $j=1$ can be omitted, since the latter ones do not
contribute to the AMM.

As far as the calculation of the electromagnetic
form factors is concerned, the change of the order of PV mass
limits also brings some new features to the procedure.
Eq.~(\ref{EMVn}), and following from it Eq.~(\ref{e0n}), are not
anymore valid. The same relates to Eq.~(\ref{Jb})
which is a particular case of Eq.~(\ref{EMVn}). As a result, we
can not rely on the renormalized expression~(\ref{EMVgf}). Indeed, the norm of
the two-body sector $I_2^{(2)}$
in the two-body approximation was calculated in
Ref.~\cite{kms_08} [see Eq.~(49) there]. Amputating from it the
factor $g^2$, we find
\begin{eqnarray}
\bar{I}_2^{(2)}&=&\frac{1}{8\pi^2}\int_0^{\infty}dR_{\perp}\,R_{\perp}
\int_0^1dx\,x\sum_{i,j=0}^1(-1)^{i+j}\nonumber\\
 &&\times\frac{R_{\perp}^2+[(1-x)m+m_i]^2}{[R_{\perp}^2+(1-x)\mu_j^2+
 xm_i^2-x(1-x)m^2]^2}.\nonumber\\
&&\label{I22}
\end{eqnarray}
Whereas, for the
electromagnetic current
$\bar{J}_2^{(2)}(0)\equiv J^{(2)}_{2}(0)/g^2$,
where $J^{(2)}_{2}$ is the two-body component of $J^{(2)}$ defined
by Eq.~(\ref{Jph}) with the EMV $G^{\rho}$ found in the
two-body approximation, the calculation
gives (the details can be found in Ref.~\cite{kms_10}):
\begin{widetext}
\begin{eqnarray}
\label{J220}
\bar{J}_2^{(2)}(0)&=&\frac{1}{8\pi^2}\int_0^{\infty}dR_{\perp}\,R_{\perp}
\int_0^1dx\,x\sum_{i,i',j=0}^1(-1)^{i+i'+j}\nonumber\\
&&\times\frac{R_{\perp}^2+(1-x)^2m^2+(1-x)m(m_i+m_{i'})
+m_im_{i'}}{[R_{\perp}^2+(1-x)\mu_j^2+
 xm_i^2-x(1-x)m^2][R_{\perp}^2+(1-x)\mu_j^2+
 xm_{i'}^2-x(1-x)m^2]}.
\end{eqnarray}
\end{widetext}
Both integrals~(\ref{I22})
(the two-body norm)
and~(\ref{J220}) (the two-body current
at zero momentum transfer) converge in the limits
$m_1\to\infty$ for finite $\mu_1$ and $\mu_1\to\infty$ for finite
$m_1$. As mentioned above, in the limit $m_1\to\infty$ for finite
$\mu_1$, they coincide with each other and satisfy the relation
(\ref{Jb}). Taking the opposite limit $\mu_1\to\infty$ for finite
$m_1$  means retaining only the terms with $j=0$ (the physical boson
index)
in the sums over $j$. One can easily check that in this limit the
difference $\bar{J}_2^{(2)}(0)-\bar{I}_2^{(2)}$ is not zero. For large values of $m_1$, it tends to a finite, mass independent, value:
\begin{equation}
\bar{J}^{(2)}_2(0)-\bar{I}_2^{(2)}=\frac{1}{16\pi^2}+O\left(\frac{m}{m_1}
\log\frac{m_1}{m}\right),\label{I22J22}
\end{equation}
while both $\bar{J}_2^{(2)}(0)$ and $\bar{I}_2^{(2)}$ taken
separately diverge as
$\log(m_1/m)$, when $m_1\to\infty$. So, under the PV regularization scheme, the
contribution of a particular Fock sector to the charge form factor
at zero momentum transfer may not coincide with the norm of this
sector.

To get a renormalized expression for $J(Q)$, we repeat the same
steps as in Sec.~\ref{elmfac}, but without using Eqs.~(\ref{e0n})
and~(\ref{Jb}). From the renormalization condition~(\ref{Jrenc})
considered, by turn, in the two- and three-body truncated Fock
spaces, we determine the bare electromagnetic coupling constants
$e_{02}$ and $e_{03}$.
Now both of them differ from the physical charge $e$.
Substituting them into the general formula~(\ref{EMV1}), we arrive
at the result for the renormalized $J(Q)$, which differs from
Eq.~(\ref{EMVgf}) by the substitution $\bar{I}_2^{(2)}\to
\bar{J}_2^{(2)}(0)$.

We compare in Table~\ref{table} the numerical results
for the AMM, obtained in this way, with those
found in Sec.~\ref{yukawa} in the limit $m_1\to\infty$.
The AMM is considered as a function of the PV mass
which is kept finite ($m_1$, if the limit $\mu_1\to\infty$ has been taken, and vice versa).
For convenience of the comparison, we took the same sets of finite PV mass values.

\begin{table} [btph]
\caption{The
anomalous magnetic moment calculated for $\alpha=0.8$
in the two different limits of the PV masses.\\}\label{table}
\begin{tabular}{|c|c|c|} \hline
PV mass kept & AMM when
& AMM when\\
finite ($\mu_1$ or $m_1$)& $m_1\to\infty$ & $\mu_1\to\infty$\\
\hline
5 & 0.1549 & 0.1454\\
\hline
10 & 0.1641 & 0.1630\\
\hline
25 & 0.1690 & 0.1704\\
\hline
50 & 0.1702 & 0.1715\\
\hline
100 & 0.1706 & 0.1716\\
\hline
250 & 0.1708 & 0.1714\\
\hline
500 & 0.1709 & 0.1713\\
\hline
\end{tabular}
\end{table}

If each of the finite PV masses is much larger than all physical
masses, the values of the AMM, obtained in both limits, coincide
within the computational accuracy (about $0.2\%$), as it should be if
the renormalization procedure works properly. We can thus choose
any convenient order of the infinite PV mass limits. Since the
equations for the Fock components are technically simpler in the
limit $m_1\to\infty$, we continue working with the vertex
functions and the EMV taken in this limit. The independence of
physical results on the order in which the
infinite PV mass limit is taken and, hence, on the way we use to
get rid of the bare parameters, is a strong evidence of the
self-consistency of our renormalization scheme.

%%%%%%%%%%%%%%%%%%%%%%%%%%%%%%%%%%%%%%%%%%%%%%%%
\section{Antiparticle degrees of freedom} \label{anti}
\subsection{Contribution to the two-body vertex function}
We can extend the Fock
decomposition of the
fermion state vector by introducing the antifermion d.o.f.
In the lowest (also three-body) approximation
this corresponds to adding the $ff\bar{f}$ Fock sector to those
previously introduced ($f$, $fb$, and $fbb$).
We have already considered the role of the three-body Fock sector
with an antiparticle within the pure scalar model (a heavy scalar
boson interacting with light scalar bosons) in Ref.~\cite{mstk}.
We perform here a similar study in the Yukawa model.

\begin{figure}[bth]
\includegraphics[width=8pc]{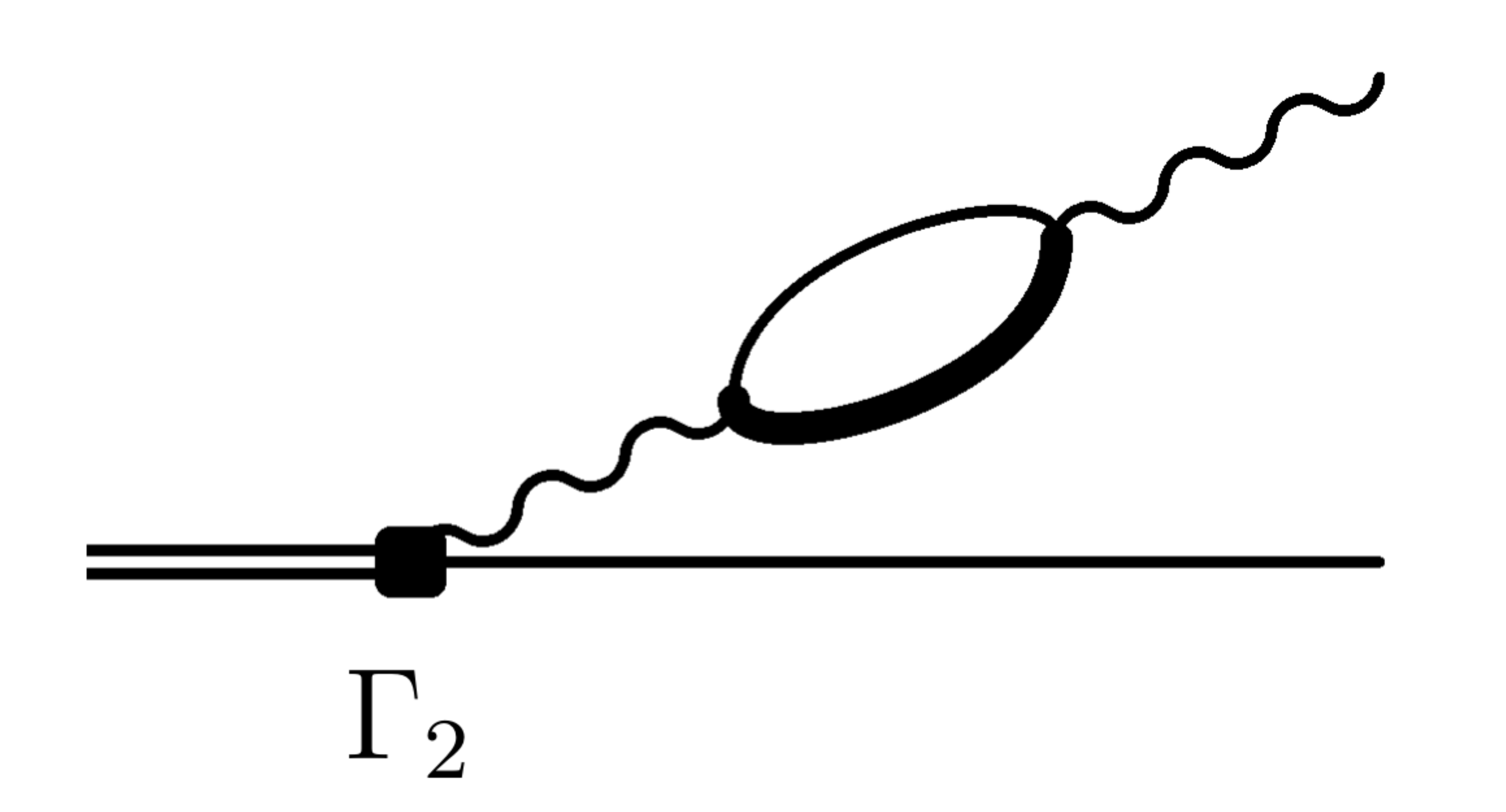}
\caption {Contribution of the $ff\bar{f}$ Fock
sector to the two-body vertex: polarization correction
to the boson line. The antifermion is
shown by the thick  line.
}\label{BbSE}
\end{figure}

The
antifermion d.o.f. contributions to the two-body
vertex are of the following two
types:

({\it i}) The first one corresponds to standard
fermion-antifermion polarization corrections to a boson
line, as shown in Fig.~\ref{BbSE}.

\begin{figure}[bth]
\includegraphics[width=18pc]{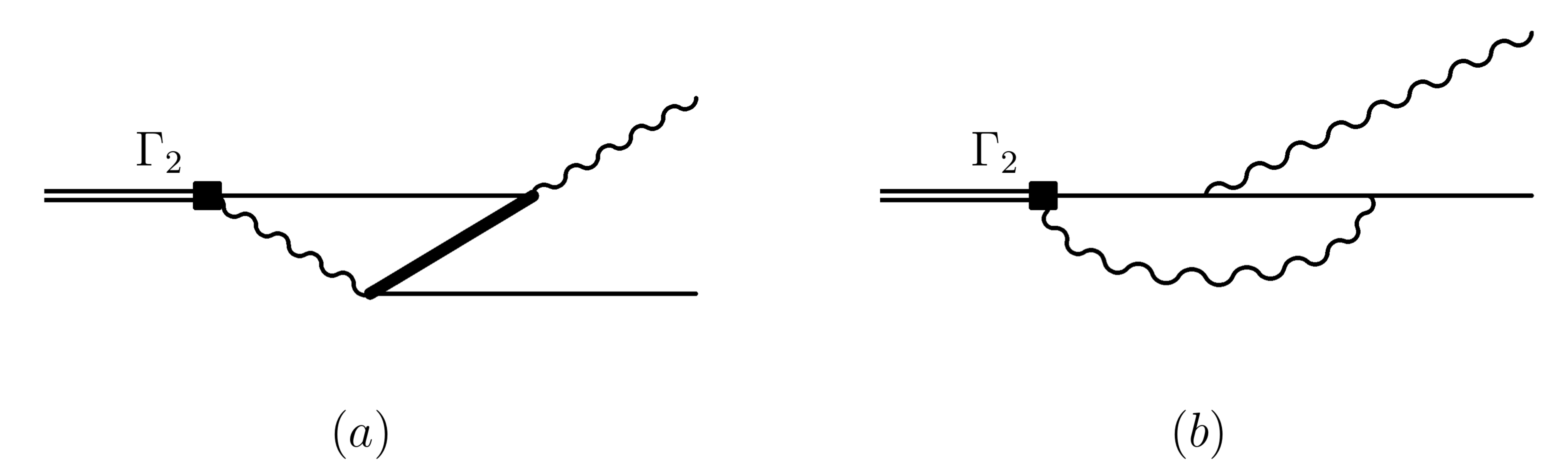}
\caption{Contributions of
the $ff\bar{f}$ (a) and $fbb$ (b) Fock
sectors to the two-body vertex due to the
transition amplitude $fb\to fb$.
}\label{antifig}
\end{figure}

({\it ii}) The second one corresponds to transition amplitudes
$fb\to fb$
associated to the excitation of antifermion d.o.f. from
a
fermion line, as shown in
Fig.~\ref{antifig}(a).

These two contributions ({\it i}) and ({\it ii})
have different nature. The first one is the $f\bar{f}$ loop. The
second contribution is the fermion-boson loop. In addition, the
diagram in Fig.~\ref{antifig}(a), by changing the order of vertices with
respect to the light-front time, evolves to the $fbb$ Fock sector
contribution, as shown in Fig.~\ref{antifig}(b). We shall
consider, in this paper,  the contributions of the second kind
only, which corresponds to the quenched approximation in LFD.

\begin{figure}[bth]
\includegraphics[width=20pc]{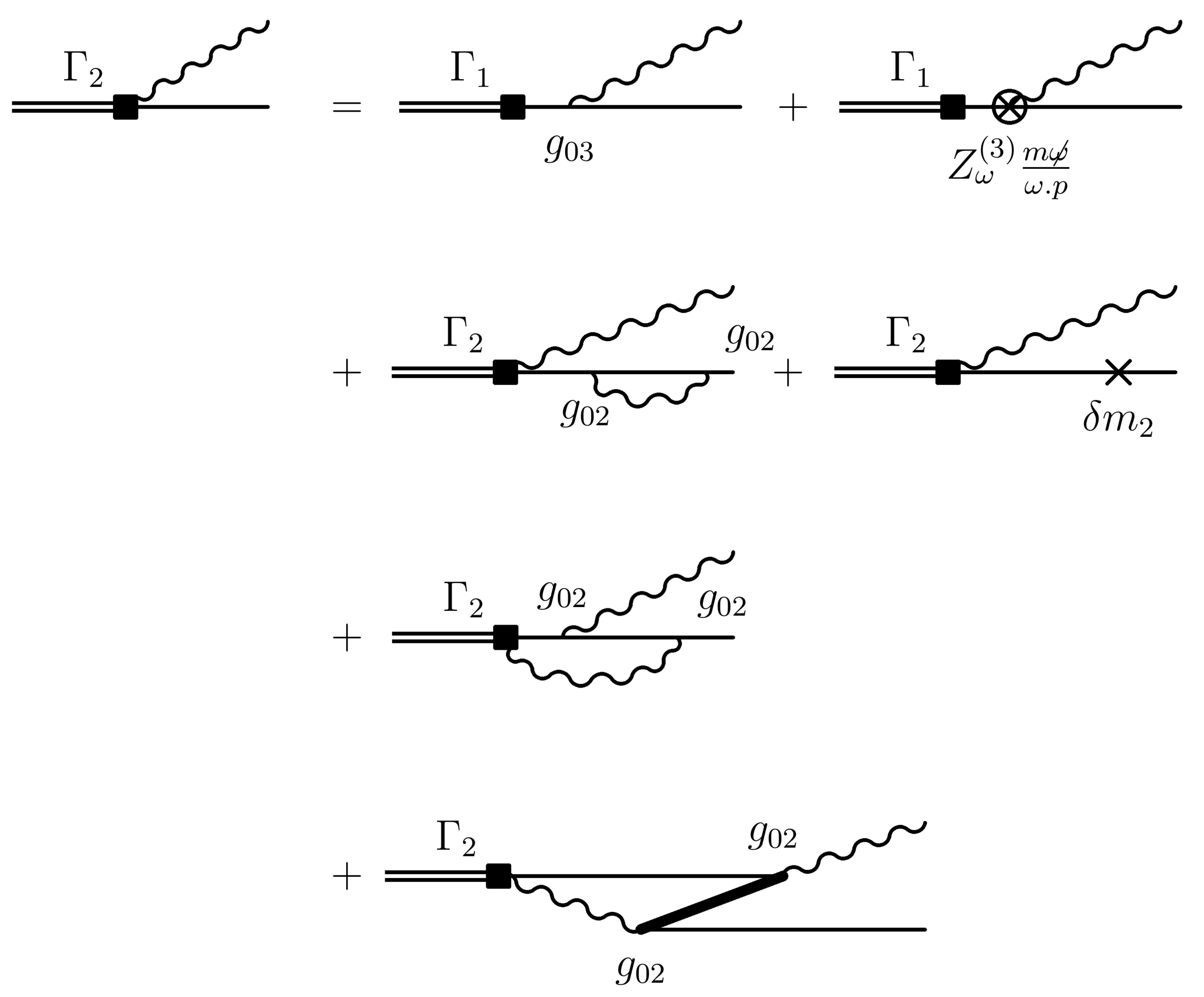}
\caption {Graphical representation of the equation for the
two-body vertex function $\Gamma_2^{(3)}$ including the
contribution of antifermion d.o.f. in the quenched approximation.}\label{fffbar}
\end{figure}

The system of equations for the Fock components can be obtained by
direct generalization of the procedure exposed in Ref.~\cite{mstk}
for the scalar case. We introduce one more three-body
($ff\bar{f}$) Fock component, in addition to the $f$, $fb$, and
$fbb$ ones considered in Sec.~\ref{ffbfbb}. In the three-body
approximation, this new Fock component is easily expressed through
the two-body component, as well as the $fbb$ one. As a result, we
obtain a closed (matrix) equation for the two-body vertex
function, as given, in the quenched approximation, by
Fig.~\ref{fffbar}. It differs from the equation in the $f+fb+fbb$
approximation, shown in Fig.~\ref{eqgamma2}, by
an additional term on the right-hand side (the last diagram
in Fig.~\ref{fffbar}).

We should pay attention to the fact of using the same
constant, $g_{02}$, in the elementary vertices $f\leftrightarrow
f+b$ and $b\leftrightarrow f+\bar{f}$. Strictly speaking, this is
not mandatory from the point of view of the general FSDR rules,
because contributions from $fbb$ and $ff\bar{f}$ states represent
different Fock sectors. Nevertheless, we assign to these vertices
the same factor $g_{02}$, which seems quite natural in this first study of the influence of antifermion d.o.f.

Since the general structure~(\ref{g2}) of the two-body vertex function
in the Yukawa model is universal,
and since we do not consider in this study polarization corrections to boson lines, the renormalization
condition~(\ref{b1oms}) is untouched. The second condition~(\ref{b2oms}) also does not change, because it is universal.
We can thus proceed further in the same way as in Sec.~\ref{eqfock}. The form of the system of equations
for the functions  $h_i^j$ and $H_i^j$
remains the same as in Eqs.~(\ref{hHm1}), but
the
integral terms $i_0^j$ and $I_0^j$ [see Eqs.~(\ref{inIn})]
obtain additive $j$-independent contributions according to
\begin{equation}
\label{ibar}
i_{0}^j\to i_{0}^j+\bar i_{0},\,\,\,\,\,\,\,\, I_{0}^j\to
I_{0}^j+\bar I_{0},
\end{equation}
where
\begin{subequations} \label{cbar}
\begin{eqnarray}
\bar{i}_{0}(R_{\perp},x)& =&\int_{0}^{\infty} R'_{\perp}d R'_{\perp}\int_{1-x}^1 dx'
\sum_{j'=0}^1(-1)^{j'}\nonumber\\
&&\times
\left[\bar{c}_{0}h_0^{j'}(R'_{\perp},x')+
\bar{C}_{0}H_0^{j'}(R'_{\perp},x')\right],\nonumber \\
&&\label{inbar}\\
\bar{I}_{0}(R_{\perp},x)& =&\int_{0}^{\infty} R'_{\perp}d R'_{\perp}\int_{1-x}^1 dx'
\sum_{j'=0}^1(-1)^{j'}\nonumber\\
&&\times
\left[\bar{c}'_{0}h_0^{j'}(R'_{\perp},x')+\bar{C}'_{0}H_0^{j'}(R'_{\perp},x')\right].
\nonumber\\
&& \label{Inbar}
\end{eqnarray}
\end{subequations}
The integral terms $i_1^j$ and $I_1^j$ do not change.
The coefficients $\bar{c}_0$, $\bar{C}_0$, $\bar{c}'_0$, and $\bar{C}'_0$
determining antifermion contributions are given
in Appendix~\ref{ap1}. Note that the limits of the integration
over $dx'$ in Eqs.~(\ref{cbar}) differ from those in Eqs.~(\ref{inIn}).

%%%%%%%%%%%%%%%%%%%%%%%%%%%%%%%%%%%%
\subsection{Numerical results}
\label{antinum}
We have solved numerically the system of equations~(\ref{hHm1})
with the integral terms modified according to Eq.~(\ref{ibar}),
for the same set of parameters as in Sec.~\ref{yukawa}: $m=0.938$,
$\mu=0.138$, and $\alpha=0.5$, $0.8$,  and $1.0$. Along with the
functions $h_i^j$ and $H_i^j$, we calculate also the bare
parameters $g_{03}'$ and $Z'_{\omega}$ defined by
Eqs.~(\ref{eqg03p}) and~(\ref{eqZomp}), respectively. In truncated
Fock space, both of them are functions of $x$, according to
Eqs.~(\ref{barex}), where the integral terms
include now antifermion contributions. Besides that, they depend
also on the PV mass $\mu_1$.

\begin{figure}[bth]
\begin{center}
\includegraphics[width=8.5cm]{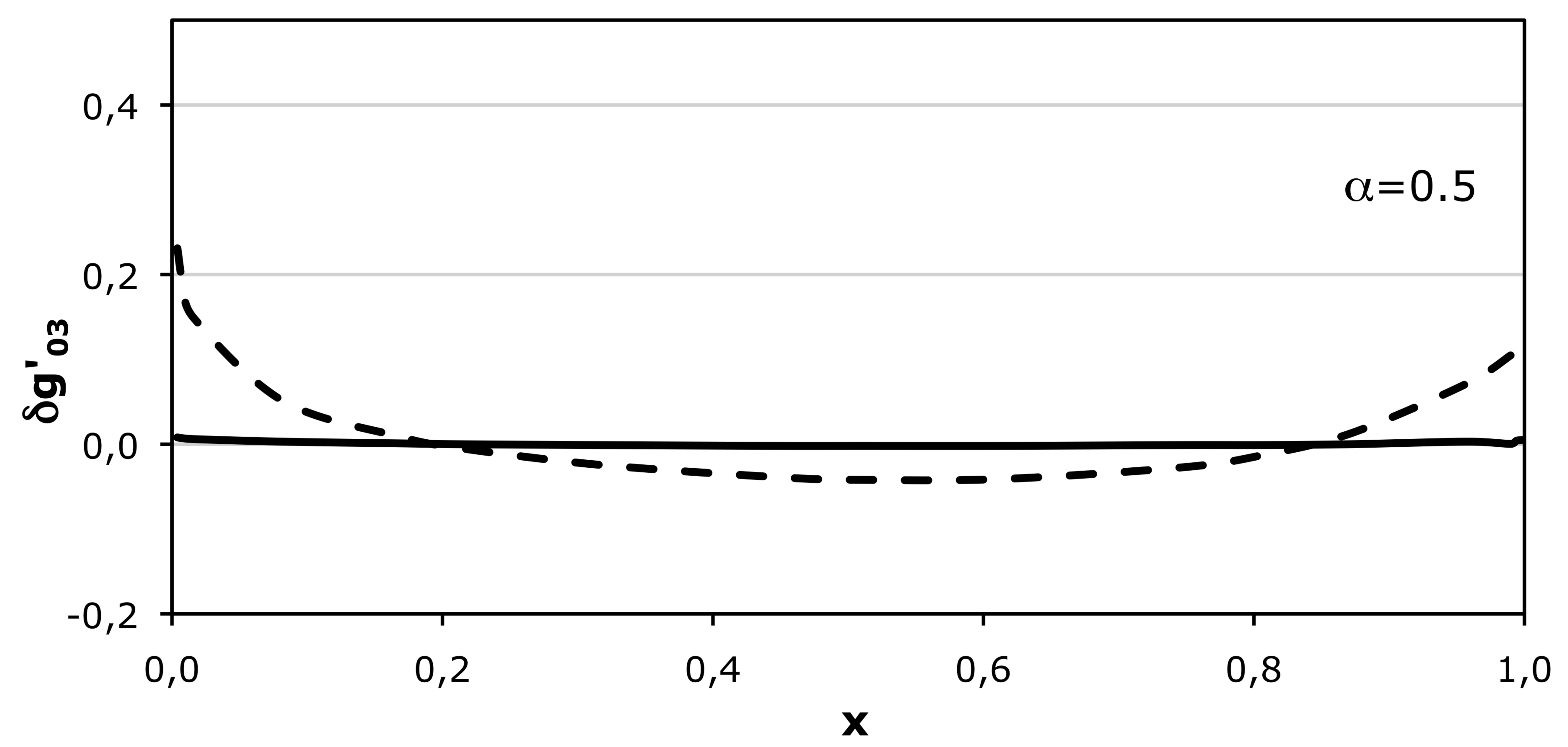}
\includegraphics[width=8.5cm]{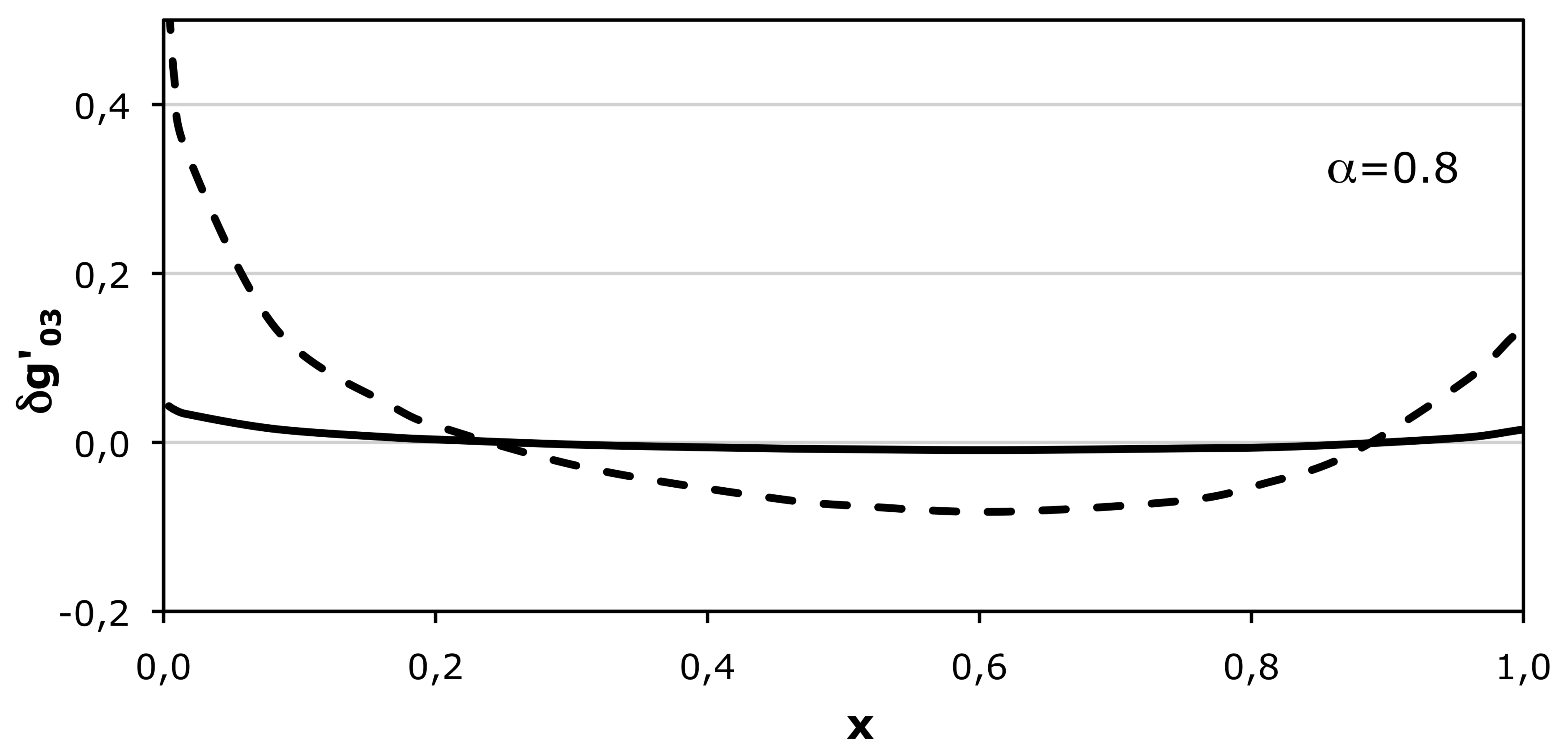}
\includegraphics[width=8.5cm]{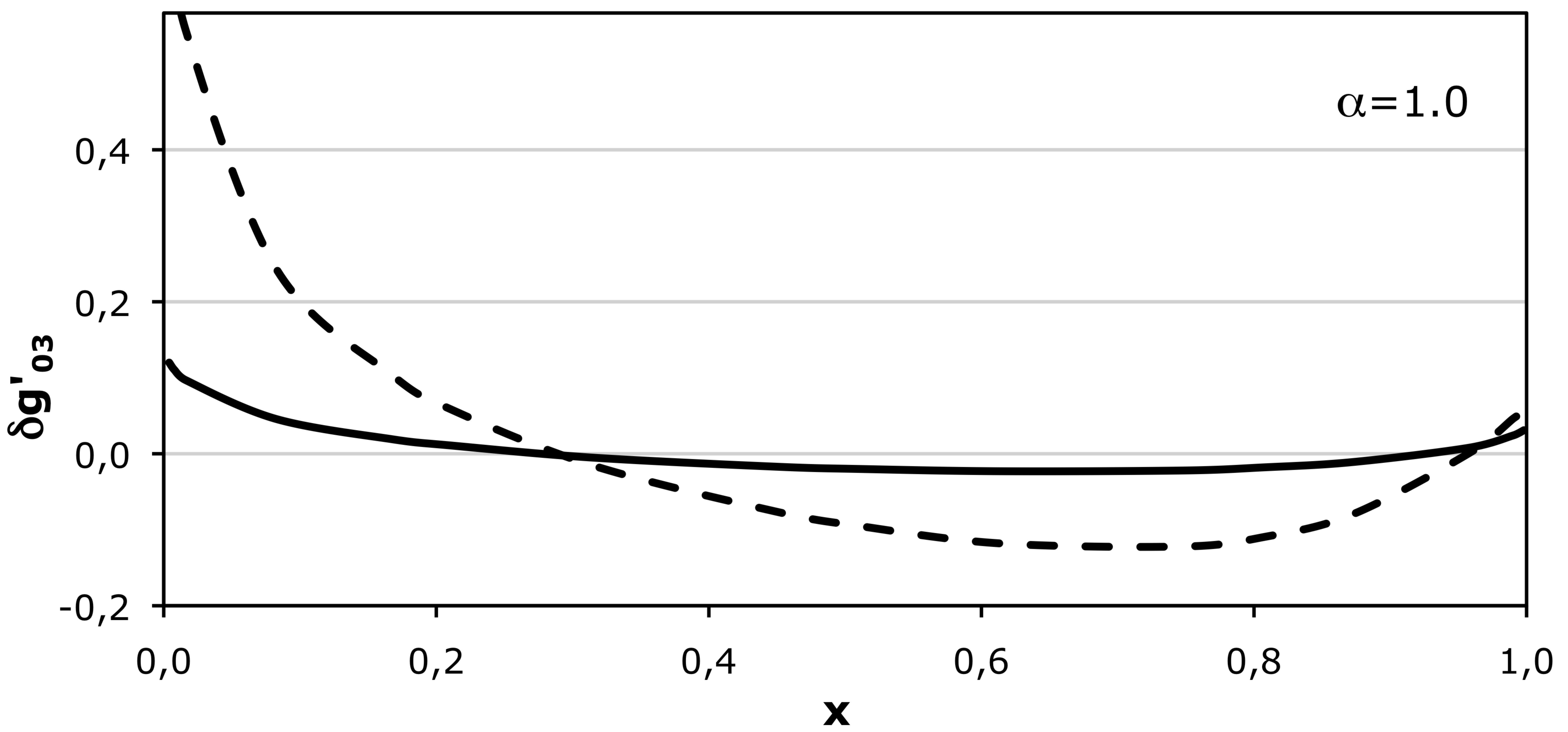}
\end{center}
\caption{$x$-dependence of the bare coupling constant
$g'_{03}$, calculated relatively to its mean value over the interval
$x\in[0,1]$, for $\alpha =0.5$ (upper plot),
$\alpha = 0.8$ (middle plot) and $\alpha = 1.0$ (lower plot),
calculated for $\mu_1=100$. The solid (dashed) lines correspond
to the results obtained with (without) the $ff\bar{f}$ Fock sector
contribution.}\label{xdepg}
\end{figure}

\begin{figure}[bth]
\begin{center}
\includegraphics[width=8.5cm]{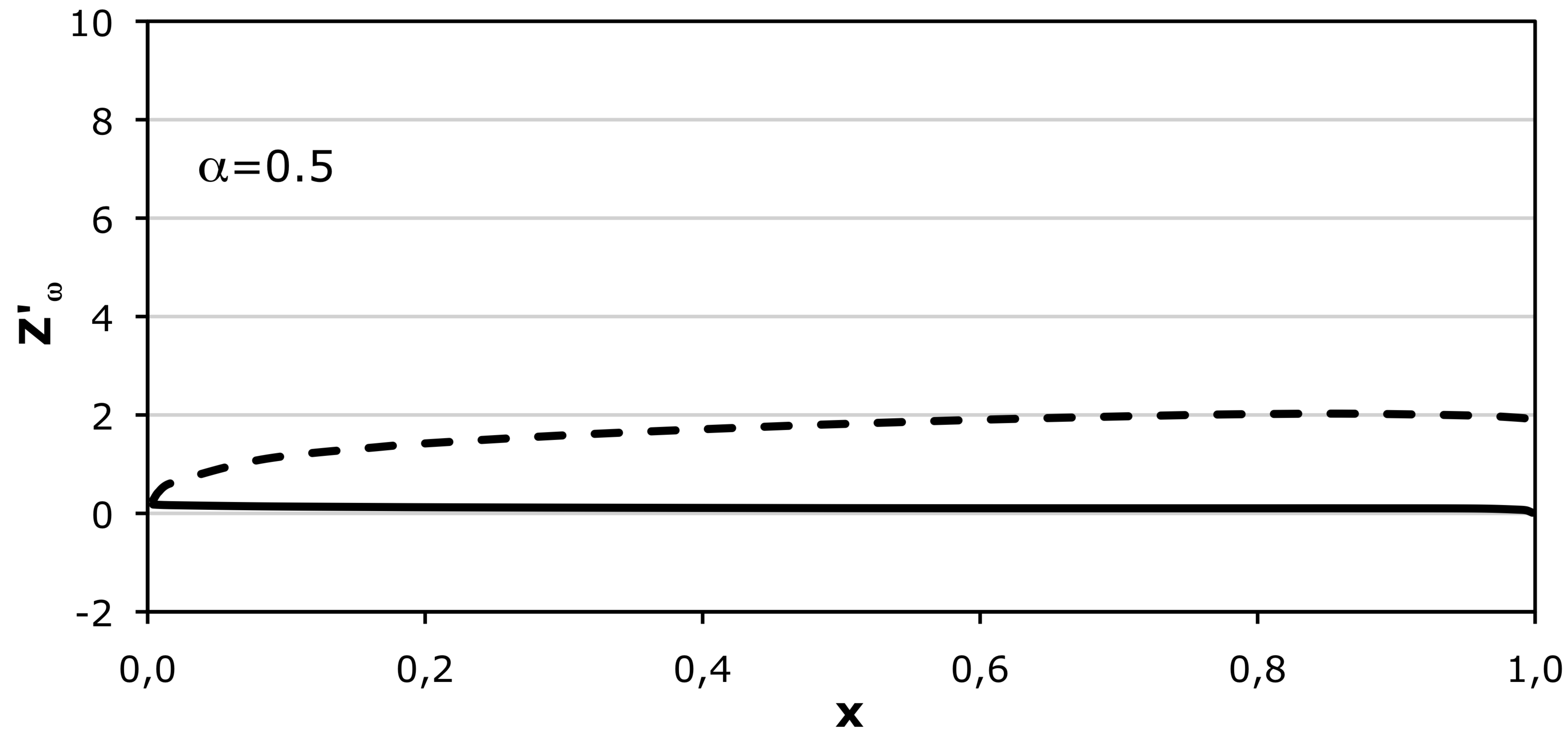}
\includegraphics[width=8.5cm]{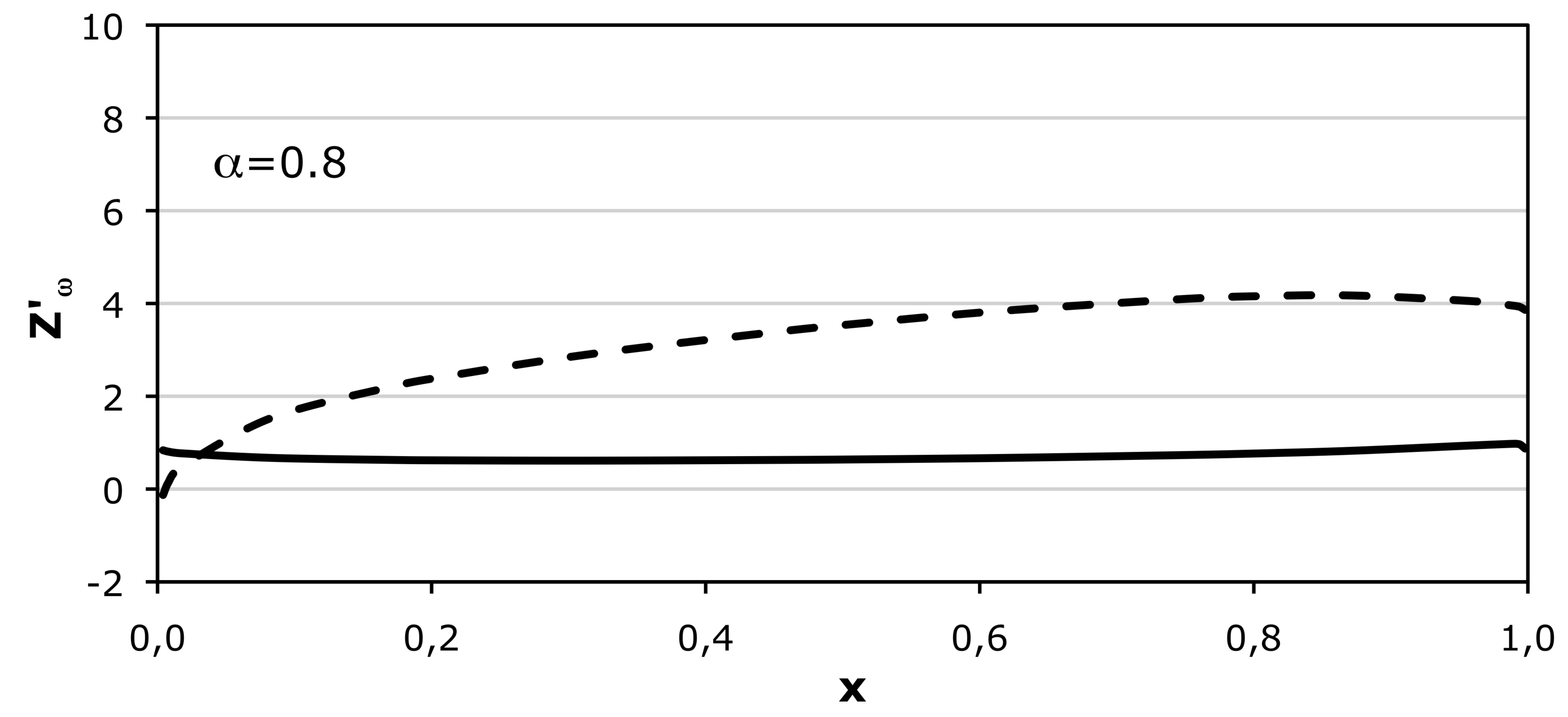}
\includegraphics[width=8.5cm]{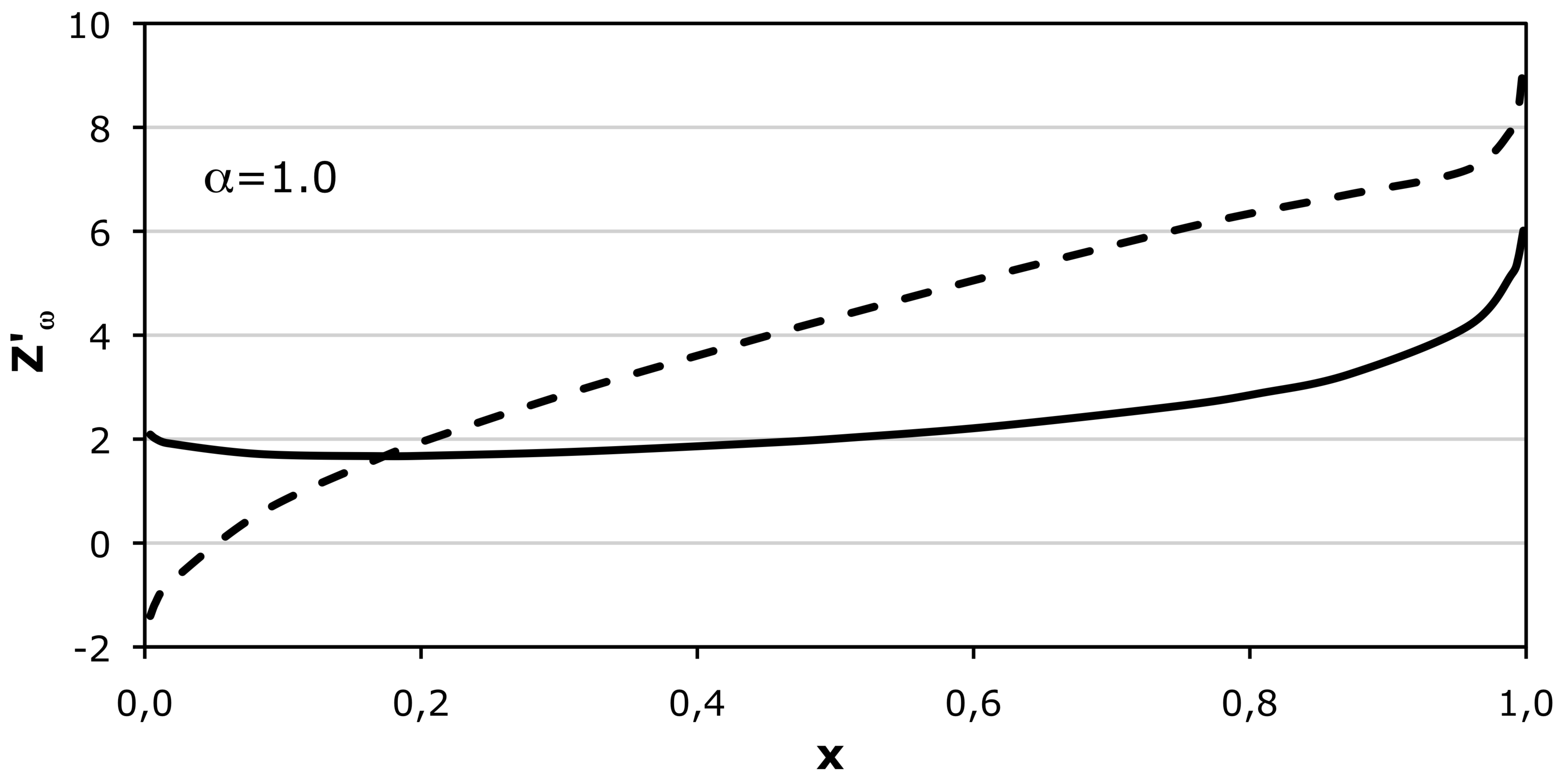}
\end{center}
\caption{$x$-dependence of the counterterm
$Z'_{\omega}$ for $\alpha =0.5$ (upper plot), $\alpha = 0.8$ (middle
plot), and $\alpha = 1.0$ (lower plot), calculated for $\mu_1=100$. The solid (dashed) lines
correspond to the results obtained with (without) the $ff\bar{f}$
Fock sector contribution.}\label{xdepZ}
\end{figure}

We plot in Figs.~\ref{xdepg} and~\ref{xdepZ}  these bare
parameters as a function of $x$, each for $\alpha=0.5$, $0.8$, and
$1.0$, at a typical value $\mu_1=100$. In Fig.~\ref{xdepg}
the relative  value of $g'_{03}$ with respect to its mean value
$\bar{g'}_{03}$ over the interval $0 \leq x \leq
1$ is shown, i.e. we plot the quantity
$$
\delta g'_{03}(x)= [g'_{03}(x)-\bar g '_{03}]/\bar g'_{03},
$$
where $\bar{g}'_{03}=\int_0^1g'_{03}(x)dx$.
For comparison, we show also on these plots the same functions
calculated without antifermion contributions. The most interesting
fact is that the function $g'_{03}(x)$, which exhibits strong
$x$-dependence in the $f+fb+fbb$ approximation, becomes almost a
constant, if the $ff\bar{f}$ Fock sector is included. Concerning
the function $Z'_{\omega}(x)$, it shows a similar tendency as well, with a bit
stronger $x$-dependency than $g'_{03}(x)$. In
addition, the magnitude of $Z'_{\omega}(x)$ is
reasonably smaller than that calculated in the $f+fb+fbb$ truncated Fock space.

The fact that $g'_{03}(x)$ and $Z'_{\omega}(x)$ are close to
constants is not a specific property of the Yukawa model, since we
already encountered similar features in our studies of the scalar
model~\cite{mstk}. In contrast to the latter, where we
incorporated particle-antiparticle loop contributions as well, we
reproduced the property $g'_{03}(x)\approx \mbox{const}$ in the
Yukawa model within the quenched approximation. From here it follows that
namely the contribution shown in Fig.~\ref{antifig}(a)
is responsible
for this property. This was the main reason of using here the
quenched approximation which allows us to keep such an important
property of the bare parameters, on the one hand, and to retain
the renormalization condition~(\ref{b1oms}) and avoid
complications connected with the fermion-antifermion loop
renormalization, on the other hand.

For a deeper understanding of what happens with the
bare parameters, when the antifermion d.o.f. are involved, let
us consider the following simple approximation. It is instructive to solve the equation
in Fig.~\ref{fffbar} by iterations, taking, as the zero order
approximation, the value of the two-body vertex function obtained
for the $f+fb$ Fock space truncation, i.~e. with $\Gamma_2=g$. We then
substitute this value on the right-hand side of this equation and
calculate the first iteration, $\Gamma_2(R_{\perp},x)$, in the
limit $m_1\to\infty$. We can then calculate  $\Gamma_2(R^*_{\perp}(x),x)$
[see Eq.~(\ref{R*})],
with both constituent particle legs corresponding to the physical
particles ($i=j=0$). This  just determines the functions
$b_{1,2}(R^*_{\perp}(x),x)$ which enter into the renormalization
conditions~(\ref{b2oms}) and~(\ref{b1oms}). Since we are
interested in the $x$-dependence of these functions, it is enough
to calculate the contributions to the latter ones from the
diagrams shown in Fig.~\ref{antifig}.
By direct calculation, we find that the sum of these two
contributions to the function $b_1$ (denoted by $b_1^a$ and $b_1^b$) on the energy shell does not
depend on $x$:
\begin{equation}
\label{sumb1}
b_1^{a}(R^*_{\perp}(x),x)+
b_1^{b}(R^*_{\perp}(x),x)=\mbox{const},
\end{equation}
while the total value of $b_2$ is zero:
\begin{equation}
\label{sumb2}
b_2^{a}(R^*_{\perp}(x),x)+
b_2^{b}(R^*_{\perp}(x),x)=0.
\end{equation}
So, on the level of the first iteration, we can meet the
renormalization condition~(\ref{b1oms}) exactly, with $g'_{03}$
independent of $x$. The other condition~(\ref{b2oms}) is satisfied
automatically, without the need of an additional counterterm~(\ref{Zomega}), i.e.  $Z_{\omega}(x)=0$.
Note that the counterterm $Z'_{\omega}$ defined by Eq.~(\ref{eqZomp}) does not
turn into zero, because it includes, beside $Z_{\omega}$, the contribution from the one-body state.
We can assert only that in the approximation discussed above it becomes a constant.

This result has a simple
explanation. The diagrams shown in Fig.~\ref{antifig} with a
constant internal two-body vertex $\Gamma_2=g$ coincide with the light-front perturbative
ones taken in the order $g^3$. Since there are no other
$g^3$-order perturbative contributions to the two-body vertex
function, the sum of their amplitudes, on the energy shell, is
identical to the corresponding on-mass-shell Feynman amplitude
which is a constant and does not depend on $\omega$.
Eqs.~(\ref{sumb1}) and~(\ref{sumb2}) are direct consequences of
this fact.

We may continue iterating the equation in
Fig.~\ref{fffbar} and represent each of the functions $b_{1,2}$ as
a series in powers of the coupling constant. But these expansions,
starting with the order $g^5$, differ from the perturbative ones.
Indeed, perturbative contributions involve Fock sectors with
arbitrary number of particles (of course, only those which are
compatible with the order of perturbation), while the
nonperturbative equation for $\Gamma_2$ contains no contributions
from higher than three-body Fock sectors. As a result,
$b_{1,2}(R^*_{\perp}(x),x)$ are no more constants, i.~e. they
depend on $x$. In order to enforce the fulfilment of the
renormalization conditions, we have to introduce $x$-dependent
bare parameters $g'_{03}(x)$ and $Z'_{\omega}(x)$. The same happens
in our nonperturbative calculations, where the two-body vertex
function $\Gamma_2$ is far from being a constant.
Note that if we assigned different vertex factors to the elementary vertices
$f\to f+b$ and $b\to f+\bar{f}$, we would not get the properties~(\ref{sumb1})
and~(\ref{sumb2}).

In the $f+fb+fbb$ approximation,  the $x$-dependence of
$b_{1,2}(R^*_{\perp}(x),x)$ is governed by the diagram in
Fig.~\ref{antifig}(b) only. The lowest order
iterative contribution of the latter diagram to the function
$b_2$, at $\mu_1\gg \{m,\,\mu\}$, has the form
\begin{equation}
\label{b2inf} b_2^{b}(R^*_{\perp}(x),x)=
-\frac{g^3}{4\pi^2}\log\frac{\mu_1}{\mu} + f_2(x),
\end{equation}
where $f_2(x)$ is a  function of $x$ and it does
not depend on $\mu_1$. In order to save the renormalization
condition~(\ref{b2oms}), we have to add the
structure~(\ref{Zomega}) to the interaction Hamiltonian, with the
$x$-dependent counterterm
$Z_{\omega}(x)=-b^{b}_2(R^*_{\perp}(x),x)$  which has a constant
$\mu_1$-dependent part (divergent when
$\mu_1\to\infty$) and a finite $x$-dependent part. For higher
order iterations, terms divergent
when
$\mu_1\to\infty$ appear in the $x$-dependent part of the
counterterm as well. Hence, in our nonperturbative calculations
within the $f+fb+fbb$ truncated space, we do expect strong
$x$-dependence of the bare parameters, which is confirmed by the
results represented in Figs.~\ref{xdepg} and~\ref{xdepZ}.
\begin{figure}[bth]
\includegraphics[width=20pc]{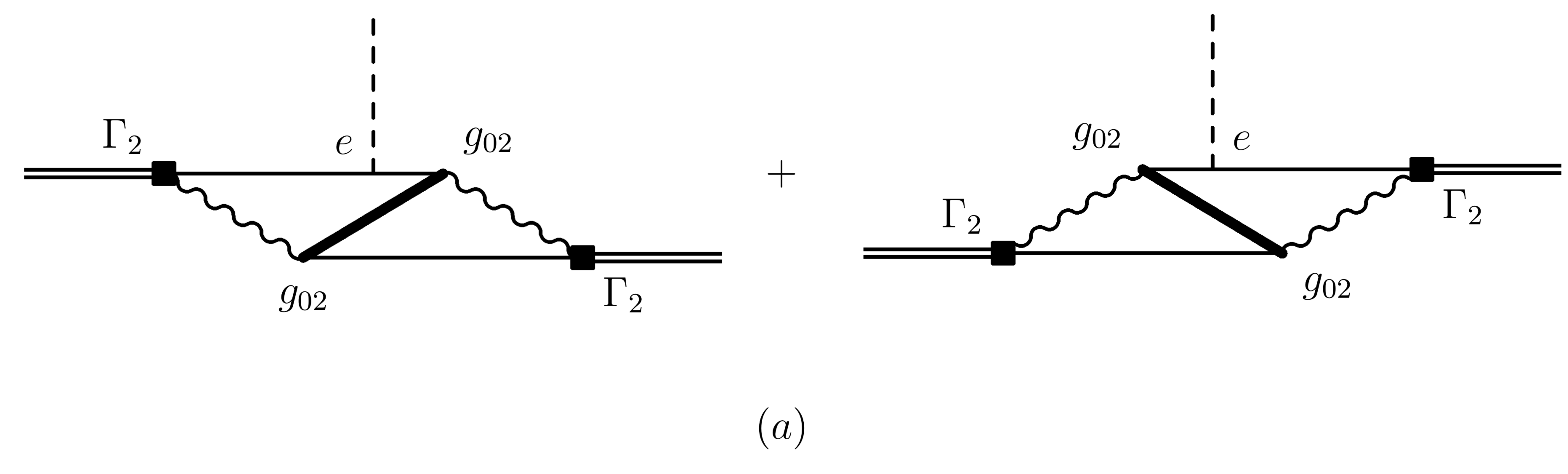}
\includegraphics[width=9pc]{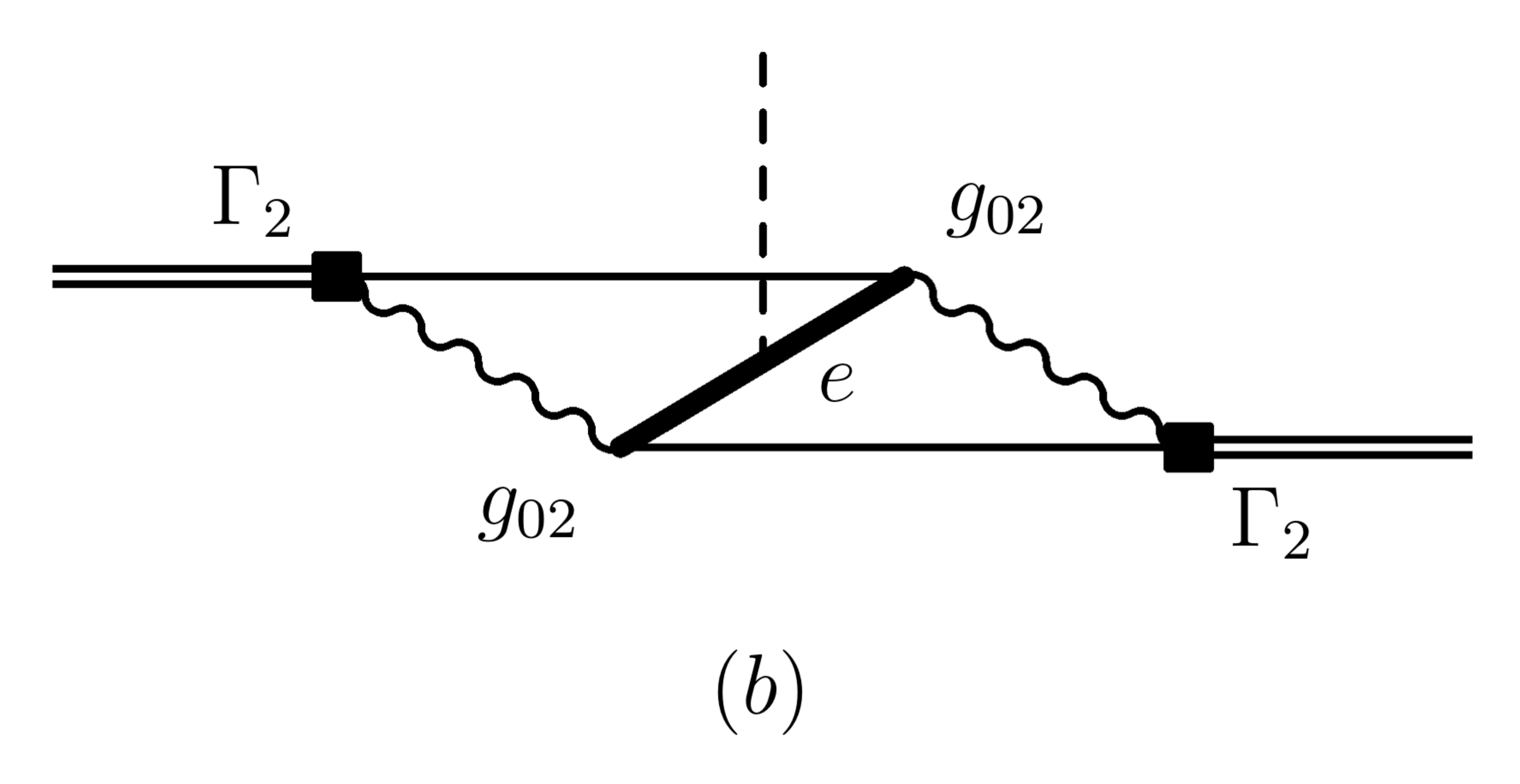}
\caption {Antifermion contributions to the electromagnetic vertex
in the quenched approximation.
}\label{elmbar}
\end{figure}

We can now  proceed to the calculation of the
electromagnetic form factors. The general method is the same as in
Sec.~\ref{elmfac}. With the antifermion d.o.f. included in the quenched approximation, one
should take into account the additional contributions to the EMV, shown
in Fig.~\ref{elmbar}. The amount of computations can be reduced by
exploiting some symmetry properties of the diagrams,  since
the two contributions in Fig.~\ref{elmbar}(a)
are exactly the same. So, we can calculate only one of
them and multiply the result by a factor of 2.
\begin{figure}[bth]
\includegraphics[width=8.5cm]{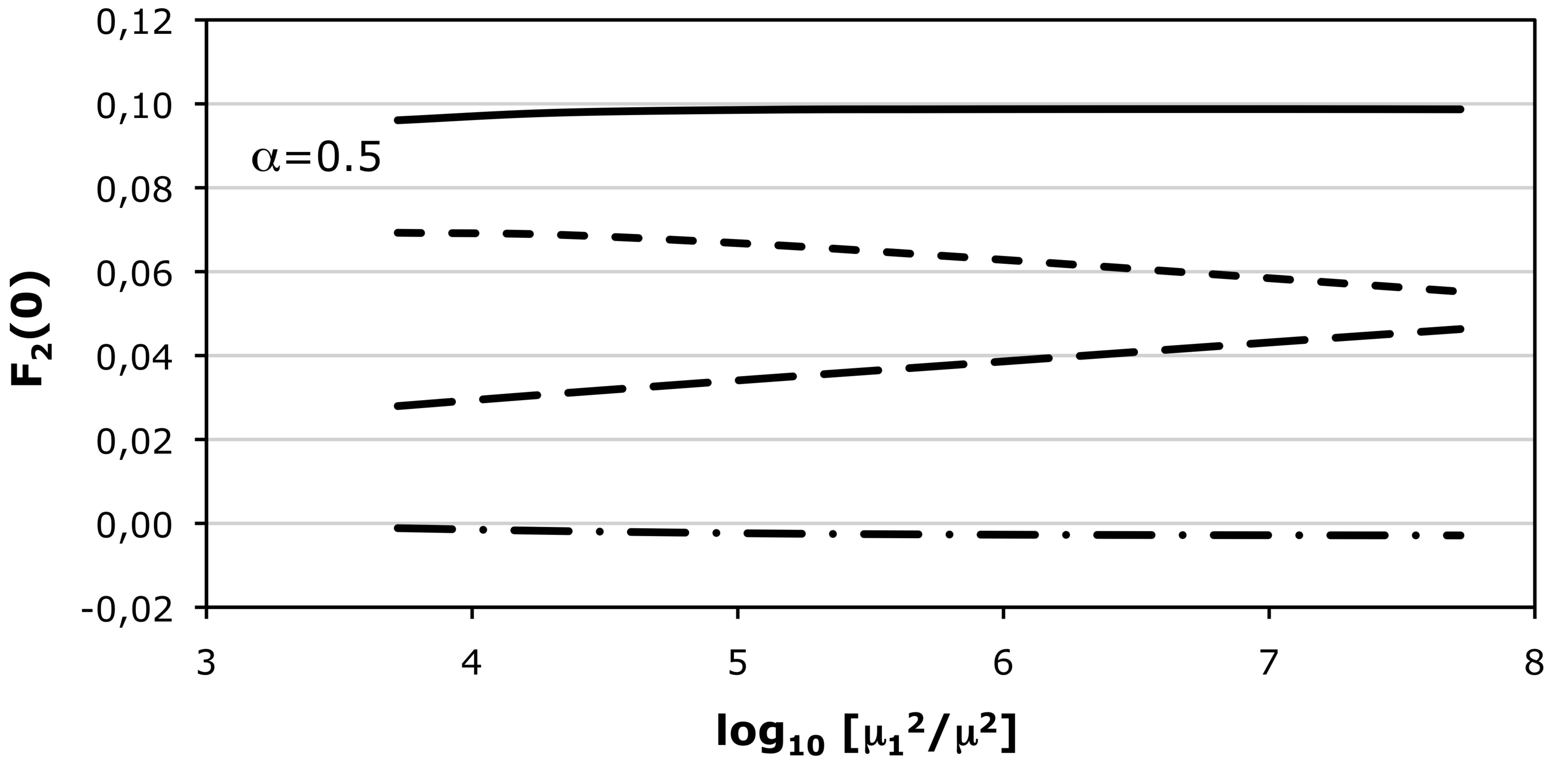}
\includegraphics[width=8.5cm]{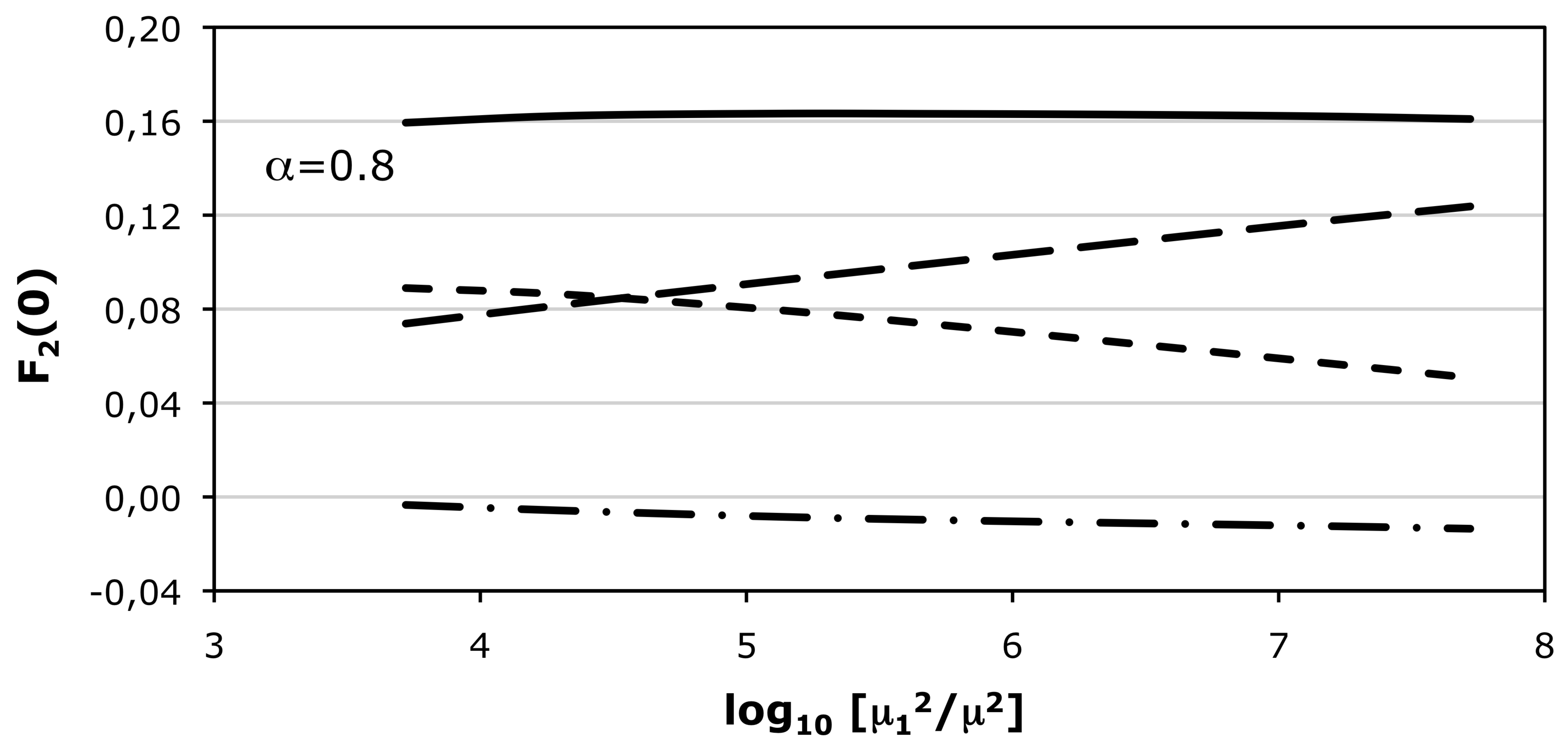}
\includegraphics[width=8.5cm]{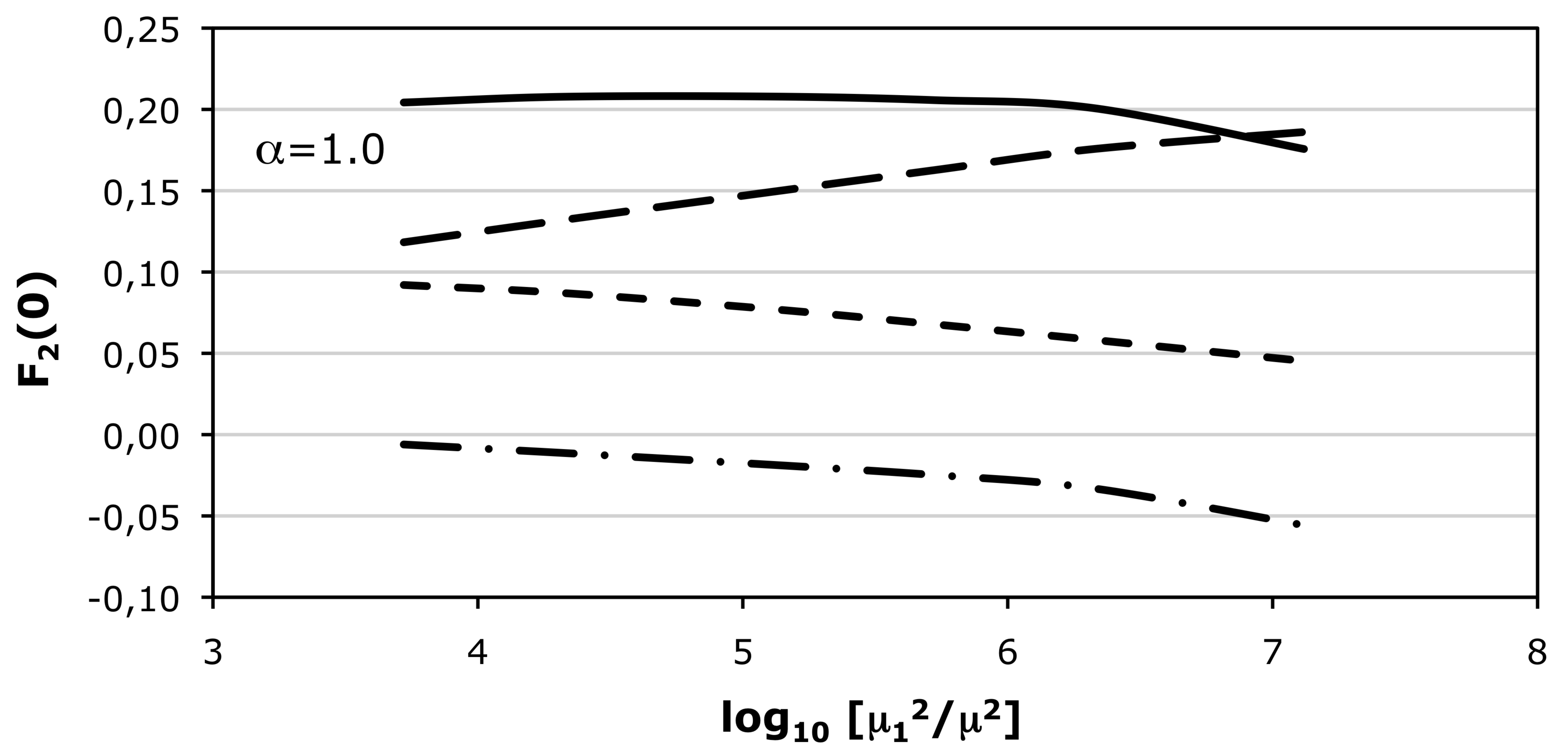}
\caption{The same as in Fig.~\ref{amm}, but including the
contributions from antifermion d.o.f. (dash-dotted
line), shown in Fig.~\ref{elmbar}.}\label{amb}
\end{figure}

For numerical calculations, we take the same values of
the particle
masses and  the coupling constant as in Sec.~\ref{yukawa}. Our results
for the fermion AMM are shown in Fig.~\ref{amb}. Although the
antifermion d.o.f. have drastic influence on the
$x$-dependence of the
bare parameters, they decrease the AMM value, as compared to that
in the $f+fb+fbb$ approximation (see Fig.~\ref{amm}), by only
$\sim 2\%$ for $\alpha=0.5$, $\sim 4\%$ for $\alpha=0.8$, and $\sim
8\%$ for $\alpha=1.0$. So small influence of the $ff\bar{f}$ Fock
sector on the AMM is caused by the smallness of the ratio
$\mu/m=0.147$. At larger $\mu$ the effect of the antifermion
d.o.f. reveals itself stronger. The AMM is stable, as a function
of $\mu_1$, if the latter is much larger than the physical masses,
but not too close to the critical mass $\mu_{1c}$ which is the same as
in the $f+fb+fbb$ approximatoin (see Sec.~\ref{yukawa}). If the latter condition
is violated, we recover the AMM instability similar to the one
obtained in the $f+fb+fbb$ truncation, with however a larger amplitude.
A visible deviation of the AMM from the constant value is
seen on the lower plot in Fig.~\ref{amb}, when $\mu_1>200$,
while $\mu_{1c}$
for $\alpha=1$ is about 1800
[$\log_{10}(\mu_{1c}^2/\mu^2)\approx 8.2$].
For $\alpha=0.5$ and $\alpha=0.8$
[$\log_{10}(\mu_{1c}^2/\mu^2)\approx 19.1$
and $\approx 11.0$, respectively] the AMM deviations from the constant in the interval
$100<\mu_1<1000$ do not exceed the computational precision level
($\sim 0.5\%$). This means that with the same accuracy we
have no any sign of uncanceled divergences.

%%%%%%%%%%%%%%%%%%%%%%%%%%%%%%%%%%%%%%%%
\section{Conclusion} \label{conc}
The results reported in this study are a first
example of a full,
nonperturbative, calculation of the properties of
relativistic compound systems
 in the FSDR framework.
The general approach is based on an expansion of
the state vector of the system considered in Fock components, within
LFD. The use of CLFD, together with an appropriate renormalization scheme
in truncated Fock space,
gives a very promising opportunity to
calculate  properties of compound systems in a regularization
scale invariant way.

The full implementation of the renormalization conditions, which
relate the on-energy-shell two-body vertex function
to physical observables,
is the last building block in our strategy to get reliable
predictions for physical
observables in the nonperturbative domain. It leads {\em a priori}, and
unambiguously, to the dependence of the bare parameters on one of
the kinematical variables of the two-body vertex function, like
for instance the longitudinal momentum fraction $x$. This
dependence must disappear in an exact calculation, i.e. when the
Fock space is not truncated.

We applied our strategy to the calculation of the
fermion electromagnetic
form factors in the Yukawa model, in the three-body Fock space
truncation. Within our numerical precision the
form factors, as a function of the momentum transfer, are
independent of the regularization scale (the PV boson
mass), as soon as the latter is large enough compared to the
typical intrinsic energy/momentum scales of the system,
but smaller than some critical value $\mu_{1c}$.

We finally investigated the role of antifermion d.o.f.
We showed that, in the leading order of perturbation theory,
the contributions to the state
vector of the Fock sector with an antifermion are precisely the ones which make the renormalization
conditions fully consistent. In that case, no extra $\omega$-dependent counterterms
are needed to restore the rotational invariance of the two-body vertex function on the energy shell,
while the $x$-dependence of the bare parameters is canceled exactly. In our nonperturbative
calculations, these contributions considerably improve
the self-consistency of the renormalization conditions.

\appendix
\section{The coefficients
$\bar{c}_0$, $\bar{C}_0$, $\bar{c}'_0$, and $\bar{C}'_0$}
\label{ap1}
It is convenient to introduce the following notations:
\begin{eqnarray*}
\eta_1&=&{R'}^2_{\perp}+x'^2m^2+(1-x')\mu^2_{j'},\\
\bar{A}&=&(1-x)x{R'}^2_{\perp}
+(1-x')x'R^2_{\perp}\\
&& +xx'(2-x-x')m^2,\\
\bar{B}&=&2(1-x)(1-x')R'_{\perp}R_{\perp},\\
\bar{D}&=&8\pi^2,
\end{eqnarray*}
and
\begin{eqnarray*}
\bar{J}_0&=&\int_0^{2\pi}\frac{d\phi'}{2\pi \bar{D}(\bar{A}+\bar{B}\cos\phi')}=
\frac{\mbox{sign}(\bar{A})}{\bar{D}\sqrt{\bar{A}^2-\bar{B}^2}},
\\
\bar{J}_1&=&\int_0^{2\pi}\frac{\cos\phi' d\phi'}{2\pi \bar{D}(\bar{A} +\bar{B}\cos\phi')}
=\frac{1}{\bar{D}\bar{B}}-\frac{\bar{A}}{\bar{B}}\bar{J}_0.
\end{eqnarray*}
Now the coefficients in Eqs.~(\ref{cbar}) take the form
\begin{eqnarray*}
\bar{c}_{0}&=&
\frac{(x-1)R_{\perp}'}{R_{\perp}\eta_1}\left[R_{\perp}R'_{\perp}\bar{J}_0\right.\\
&&\left.+m^2(xx'+2x+2x'-4)\bar{J}_1\right],\\
\bar{C}_{0}&=&\frac{(x-1)m^2}{R_{\perp}\eta_1}\left[
R_{\perp}(3x'-2)\bar{J}_0-R_{\perp}'(3x-2)\bar{J}_1\right],
\\
\bar{c}'_{0}&=&\frac{(x-1)R'_{\perp}}{\eta_1}\left[
R'_{\perp}(3x-2)\bar{J}_0-R_{\perp}(3x'-2)\bar{J}_1\right],
\\
\bar{C}'_{0}&=&\frac{(x-1)}{\eta_1}
\left[m^2(xx'+2x+2x'-4)\bar{J}_0\right.\\
&&\left.+R_{\perp}R'_{\perp}\bar{J}_1\right].
\end{eqnarray*}
Note that these coefficients do not depend on the index $j$.

%%%%%%%%%%%%%%%%%%%%%%%%%%%%%%%%%%%%%%%
\begin{acknowledgments}
Two of us (V.~A.~K. and A.~V.~S.) are sincerely grateful for the
warm hospitality of the Laboratoire de Physique Corpusculaire,
Universit\'e Blaise Pascal, in Clermont-Ferrand, where
a part of the present study was performed, under a CNRS/RAS
agreement.
\end{acknowledgments}

%%%%%%%%%%%%%%%%%%%%%%%%%%%%%%%%%%%%%%%

%%%%%%%%%%%%%%%%%%%%%%%%%%%%%%%%%%%%%%%%%%%


\begin{thebibliography}{10}

\bibitem{bckm}
D.~Bernard, Th.~Cousin, V.A.~Karmanov, and J.-F.~Mathiot, Phys.
Rev. D {\bf 65}, 025016 (2001).

\bibitem{kms_04}
 V.A.~Karmanov, J.-F.~Mathiot, and A.V.~Smirnov. Phys. Rev.
D {\bf 69}, 045009 (2004).

\bibitem{kms_07}
 V.A.~Karmanov, J.-F.~Mathiot, and A.V.~Smirnov,
Phys. Rev. D {\bf 75}, 045012 (2007).

\bibitem{kms_08}
V.A.~Karmanov, J.-F.~Mathiot, and A.V.~Smirnov, Phys. Rev. D {\bf
77}, 085028 (2008).

\bibitem{kms_10}
V.A.~Karmanov, J.-F.~Mathiot, and A.V.~Smirnov, Phys. Rev. D
{\bf 82}, 056010 (2010).

\bibitem{mstk}
J.-F.~Mathiot, A.V.~Smirnov, N.A.~Tsirova, and V.A.~Karmanov,
Few-Body Syst. {\bf 49}, 183 (2011).

\bibitem{dirac}
P.A.M.~Dirac, Rev. Mod. Phys. {\bf 21}, 392 (1949).

\bibitem{bpp}
S.J.~Brodsky, H.C.~Pauli, and S.S.~Pinsky, Phys. Rep. {\bf 301},
299 (1998).

\bibitem{karm76}
V.A.~Karmanov, Zh. Eksp. Teor. Fiz.  {\bf 71}, 399 (1976)
[Sov. Phys. JETP {\bf 44}, 210 (1976)].

\bibitem{cdkm}
J.~Carbonell, B.~Desplanques, V.A.~Karmanov, and J.-F.~Mathiot,
Phys. Rep.   {\bf 300}, 215 (1998).

\bibitem{PV}
S.J.~Brodsky, J.R.~Hiller, and G.~McCartor, Phys. Rev. D {\bf 64},
114023 (2001);\\
S.J.~Brodsky, J.R.~Hiller, and G.~McCartor, Ann. Phys. {\bf 305},
266
(2003);\\
S.J. Brodsky, J.R. Hiller and G. McCartor,  Ann. Phys. {\bf 321},
1240 (2006).

\bibitem{WG}
St.~Glazek, A.~Harindranath, S.~Pinsky, J.~Shigemitsu, and K.~Wilson,
Phys. Rev. D {\bf 47}, 1599 (1993).

\bibitem{perry}
R.J.~Perry, A.~Harindranath, and K.G.~Wilson, Phys. Rev. Lett. {\bf 65}, 2959 (1990);\\
R.J.~Perry and A.~Harindranath, Phys. Rev. D {\bf 43}, 4051 (1991).

\bibitem{hiller}
J.R.~Hiller and S.J.~Brodsky, Phys. Rev. D {\bf 59}, 016006 (1998).

\bibitem{km96}
V.A.~Karmanov and J.-F.~Mathiot, Nucl. Phys. {\bf A602}, 388
(1996).

\end{thebibliography}
\end{document}